\documentclass[11pt,a4paper]{article}
\usepackage{jstylemarg}
\usepackage[utf8]{inputenc}
\usepackage{amsmath}
\usepackage{mathtools}
\usepackage{tikz}

\newcommand{\bi}{\begin{itemize}}
\newcommand{\ei}{\end{itemize}}
\newcommand{\bea}{\begin{align}}
\newcommand{\eea}{\end{align}}
\newcommand{\be}{\begin{equation}}
\newcommand{\ee}{\end{equation}}

\newcommand{\Det}{{\text{det}}}
\newcommand{\pl}{{\partial}}

\newcommand*\circled[1]{\tikz[baseline=(char.base)]{
  \node[shape=circle,draw, inner sep=1pt] (char) {#1};}}

\newcommand{\tcb}{\textcolor{blue}}

\makeatletter
\renewcommand*\env@matrix[1][\arraystretch]{%
  \edef\arraystretch{#1}%
  \hskip -\arraycolsep
  \let\@ifnextchar\new@ifnextchar
  \array{*\c@MaxMatrixCols c}}
\makeatother

\author{Charlotte SLEIGHT}

\author{\quad Massimo TARONNA\footnote{Postdoctoral Researcher of the Fund for Scientific Research-FNRS Belgium.}}

\affiliation{Universit\'e Libre de Bruxelles
and International Solvay Institutes\\
ULB-Campus Plaine CP231, 1050 Brussels, Belgium}

\emailAdd{charlotte.sleight@gmail.com, taronnam@gmail.com}

\title{\centering
\huge{Feynman rules\\ for higher-spin gauge fields on AdS$_{d+1}$}}

\abstract{We determine the Feynman rules for the minimal type A higher-spin gauge theory on AdS$_{d+1}$ at cubic order. In particular, we establish the quantum action at cubic order in de Donder gauge, including ghosts. We also give the full de Donder gauge propagators of higher-spin gauge fields and their ghosts. This provides all ingredients needed to quantise the theory at cubic order.}

\begin{document}

\maketitle

\section{Introduction}

The Fronsdal program has been formulated to define consistent non-linear field theories which include interacting higher-spin gauge fields at the classical level \cite{PhysRevD.18.3624}. Its long standing motivation has been a deeper understanding of the symmetries behind quantum gravity. Numerous attempts have been devoted to find solutions to this problem from various perspectives, which include the Noether procedure (see \cite{Berends:1984rq,Berends:1984wp,Bengtsson:1987jt,Metsaev:2005ar,Buchbinder:2006eq,Boulanger:2006gr,Fotopoulos:2007yq,Zinoviev:2008ck,Fotopoulos:2008ka,Boulanger:2008tg,Taronna:2010qq,Manvelyan:2010wp,Manvelyan:2010jr,Sagnotti:2010at,Fotopoulos:2010ay,Manvelyan:2010je,Taronna:2011kt,Joung:2011ww,Joung:2012rv,Joung:2012fv,Campoleoni:2012hp,Joung:2012hz,Joung:2013doa,Joung:2013nma,Joung:2014aba,Sleight:2016xqq,Francia:2016weg,Buchbinder:2017nuc} and references therein, for an incomplete list of relevant works), the frame-like formalism \cite{Fradkin:1986qy,Fradkin:1987ks,Alkalaev:2002rq,Alkalaev:2010af,Vasilev:2011xf,Boulanger:2012dx,Boulanger:2013zza}, and key attempts \cite{Vasiliev:1990en,Vasiliev:2003ev,Boulanger:2011dd,Boulanger:2012bj,Boulanger:2015kfa,Bonezzi:2015igv} to obtain directly a fully non-perturbative formulation of a higher-spin theory.

However, in spite of the above remarkable efforts, all attempts are being confronted with one and the same conceptual subtlety, which is intimately related to the definition of a non-local extension of the classical field theoretic deformation problem that lies at the basis of Einstein General Relativity and QFT.

So far, it has been possible to make sense of pseudo-local higher-spin Lagrangians and equations of motion up to the cubic order, where cubic couplings have been recently fixed completely both using holography \cite{Petkou:2003zz,Bekaert:2014cea,Bekaert:2015tva,Skvortsov:2015lja,Skvortsov:2015pea,Sleight:2016dba,Sleight:2016hyl,Sleight:2017fpc} and Noether procedure \cite{Sleight:2016xqq,Sleight:2017pcz}. However, beyond the cubic order a proper extension of the functional class of local Lagrangian functionals and equations of motion is currently lacking. This goes hand in hand with the proliferation of infinitely many explicitly non-local off-shell solutions to the Noether procedure,\footnote{See e.g. \cite{Taronna:2011kt,Boulanger:2015ova,Skvortsov:2015lja,Taronna:2016ats,Taronna:2016xrm,Taronna:2017wbx,Roiban:2017iqg,Sleight:2017pcz} for a detailed discussion of this issue.} which lead to one and the same observable defined by AdS/CFT correspondence. It was further argued in \cite{Sleight:2017pcz} that no proper extension of the functional space of non-localities is possible in a properly defined generalised field theoretic context and that one may have to resort to String Theory, i.e. beyond the realm of field theory, to achieve a proper definition of higher-spin theories.

On the other hand any CFT defines, up to off-shell ambiguities, a \emph{formal} bulk \emph{field theory}. From this perspective AdS/CFT acquires a tautological meaning and can be considered as some kind of general non-local map/transform which can be inverted to fix bulk Lagrangian couplings in a process that has been referred to as holographic reconstruction \cite{Bekaert:2015tva,Bekaert:2016ezc,Sleight:2016dba,Sleight:2016hyl,Castro:2017hpx,Sleight:2017fpc,Gubser:2017tsi}. Exactly in the same way as it is possible to enlarge the functional domain to define an integral transform, the key question is about a clever choice of regularity conditions which ensure a proper independent definition of both the boundary and bulk sides of the duality.

For these reasons it is important to push beyond tree-level, and investigate quantum properties of higher-spin gauge theories independently on both the bulk and boundary sides to test the degree of non-localities. With this motivation in mind, the aim of this work is to revisit the ambient space formalism and formulate a consistent scheme to quantise higher-spin gauge theories on AdS by writing down their quantum action in a fixed gauge. Together with recently developed tools \cite{Giombi:2017hpr} to evaluate loop diagrams with external legs on AdS,\footnote{For work on quantum corrections in flat space see \cite{Ponomarev:2016jqk}. For previous investigations of quantum corrections in AdS, see \cite{Manvelyan:2004ii,Manvelyan:2008ks}.} this constitutes a key step towards quantum tests of the duality, beyond the one-loop vacuum energy checks \cite{Giombi:2013fka,Giombi:2014iua,Giombi:2014yra,Beccaria:2014xda,Basile:2014wua,Giombi:2016pvg,Pang:2016ofv,Bae:2016rgm,Bae:2016hfy,Gunaydin:2016amv,Bae:2017spv,Skvortsov:2017ldz} which only probe the free theory. In particular, in this work we write down the full cubic quantum action for the type-A theory in the de-Donder gauge and invert the corresponding kinetic terms for both ghosts and physical fields in the same gauge. We also outline a systematic procedure to formally reconstruct the full quantum action at any order in weak fields given the physical vertex for traceless and transverse fields. In the next section we provide a short summary of the main results.

\subsection{Summary of results}

In this work we determine the gauge fixed path integral of the minimal type A higher-spin gauge theory on AdS$_{d+1}$ up to cubic order fluctuations, together with associated propagators in the same gauge. 
The path integral reads
\begin{equation}
Z=\int[d\varphi][d{\bar c}][d c]\,e^{-S_q[\varphi,c,{\bar c}]}\,,
\end{equation}
with quantum action
\begin{equation}
S_q\left[\varphi,c,{\bar c}\right]=S[\varphi]+S_{\text{ghost}}\left[\varphi,{\bar c},c\right],
\end{equation}
where $\varphi$ collectively denotes the spectrum of Fronsdal fields in the type A theory, and $c$, ${\bar c}$ the corresponding ghosts. Expanding up to cubic order:
\begin{subequations}
\begin{align}
    S[\varphi]&=S^{(2)}[\varphi]+S^{(3)}[\varphi]+...\,\\
    S_{\text{ghost}}\left[\varphi,{\bar c},c\right]&=S^{(2)}_{\text{ghost}}\left[\varphi,{\bar c},c\right]+S^{(3)}_{\text{ghost}}\left[\varphi,{\bar c},c\right]+...\,.
\end{align}
\end{subequations}
The free minimal type A Fronsdal action is given by
\begin{align}
    S^{(2)}[\varphi]=\sum_{s \in 2\mathbb{N}}S^{(2)}[\varphi_s], \label{typeafree}
\end{align}
where $S^{(2)}[\varphi_s]$ is the action \eqref{Fronsdalaction} describing a free totally symmetric spin-$s$ Fronsdal gauge field $\varphi_s$ \cite{Fronsdal:1978rb}. The field $\varphi_0$ is a parity even scalar of fixed mass $m^2_0=-2\left(d-2\right)/R^2$. This is the minimal spectrum compatible with global higher-spin symmetry \cite{Fradkin:1986ka,Eastwood:2002su,Vasiliev:2003ev}.

In \S \tcb{\ref{subsec::ffcc}} we derive the cubic order action in de Donder gauge, which reads
\begin{subequations}\label{osactionint}
\begin{align}
S^{(3)}&=\sum_{s_1\geq s_2 \geq s_3} {\cal V}_{s_1,s_2,s_3}\\ \label{offshell0int}
{\cal V}_{s_1,s_2,s_3}&={\cal V}^{TT}_{s_1,s_2,s_3}+{\cal V}^{(1)}_{s_1,s_2,s_3}+{\cal V}^{(2)}_{s_1,s_2,s_3}+{\cal V}^{(3)}_{s_1,s_2,s_3},
\end{align}
\end{subequations}
where \cite{Sleight:2016dba,Sleight:2016xqq} (using the ambient space formalism reviewed in \S \tcb{\ref{subsec::revamb}}):
\begin{subequations}\label{ttactionint}
\begin{align} 
{\cal V}^{TT}_{s_1,s_2,s_3}&=f^{TT}_{s_1,s_2,s_3} \varphi_{s_1}\left(X_1,U_1\right)\varphi_{s_2}\left(X_2,U_2\right)\varphi_{s_s}\left(X_3,U_3\right)\Big|_{X_i=X}, \\ \nonumber
f^{TT}_{s_1,s_2,s_3}&= g_{s_1,s_2,s_3}{\cal Y}^{s_1}_1{\cal Y}^{s_2}_2{\cal Y}^{s_3}_3,
\end{align}
\end{subequations}
are the cubic couplings for traceless and transverse fields, with coupling constants \cite{Sleight:2016dba,Sleight:2016xqq} 
\begin{equation}\label{typeaccint}
   g_{s_1,s_2,s_3}=\frac{1}{\sqrt{N}}\frac{\pi ^{\frac{d-3}{4}}2^{\tfrac{3 d-1+s_1+s_2+s_3}{2}}}{ \Gamma (d+s_1+s_2+s_3-3)}  \prod_{i=1}^3\sqrt{\frac{\Gamma(s_i+\tfrac{d-1}{2})}{\Gamma\left(s_i+1\right)}}.
\end{equation}
Their de Donder gauge off-shell completion is given in the ambient formalism by
\begin{subequations}
\begin{multline}
{\cal V}^{(1)}_{s_1,s_2,s_3}=-\frac12\,(\pl_{{\cal Y}_1}f^{TT}_{s_1,s_2,s_3})\,U_1\cdot X_1\,\varphi_{s_1}^{\prime(1)}\varphi_{s_2}\varphi_{s_3}-\frac12\,(\pl_{{\cal Y}_2}f^{TT}_{s_1,s_2,s_3})\varphi_{s_1}\,U_2\cdot X_2\,\varphi_{s_2}^{\prime(1)}\varphi_{s_3}\\-\frac12\,(\pl_{{\cal Y}_3}f^{TT}_{s_1,s_2,s_3})\varphi_{s_1}\varphi_{s_2}\,U_3\cdot X_3\,\varphi_{s_3}^{\prime(1)}\,,
\end{multline}
\begin{multline}
    {\cal V}^{(2)}_{s_1,s_2,s_3}=\frac12\,\pl_{{\cal Y}_1}\pl_{{\cal Y}_2}f^{TT}_{s_1,s_2,s_3}\, U_1\cdot X_1\,\varphi^{\prime(1)}_{s_1}\,U_2\cdot X_2\,\varphi_{s_2}^{\prime(1)}\varphi_{s_3}+\text{cyclic}\\
    +\frac14(d-2+2{\cal Y}_2\pl_{{\cal Y}_2})\pl_{{\cal Y}_1}^2\pl_{{\cal Y}_2}f^{TT}_{s_1,s_2,s_3}\,\varphi^\prime_{s_1}\,U_2\cdot X_2\,\varphi_{s_2}^{\prime(2)}\varphi_{s_3}+\text{cyclic}\,,
\end{multline}
\begin{multline}
    {\cal V}^{(3)}_{s_1,s_2,s_3}=\frac18\,(d+2{\cal Y}_1\pl_{{\cal Y}_1})(d+2{\cal Y}_2\pl_{{\cal Y}_2})(d+2{\cal Y}_3\pl_{{\cal Y}_3})\pl_{{\cal Y}_1}^2\pl_{{\cal Y}_2}^2\pl_{{\cal Y}_3}^2f\,\varphi_{s_1}^{\prime(1)}\varphi_{s_2}^{\prime(1)}\varphi_{s_3}^{\prime(1)}\\
    +\frac18(d+2{\cal Y}_3\pl_{{\cal Y}_3})(d-2+2{\cal Y}_1\pl_{{\cal Y}_1})\pl_{{\cal Y}_1}\pl_{{\cal Y}_2}^2\pl_{{\cal Y}_3}^2\,U_1\cdot X_1\,\varphi_{s_1}^{\prime(2)}\varphi_{s_2}^\prime\varphi_{s_3}^{\prime(1)}+\text{cyclic}\\
    -\frac12\,\pl_{{\cal Y}_1}\pl_{{\cal Y}_2}\pl_{{\cal Y}_3}f\,U_1\cdot X_1\,\varphi_{s_1}^{\prime(1)}\,U_2\cdot X_2\,\varphi_{s_2}^{\prime(1)}\,U_3\cdot X_3\,\varphi_{s_3}^{\prime(1)}\,,
\end{multline}
\end{subequations}
which are fixed by requiring gauge invariance in the de-Donder gauge.

The free ghost action can be fixed by the linear gauge transformations, and is given by (see \S \tcb{\ref{subsec::gcc}} and, for notation, \S \tcb{\ref{sec::notconvamb}})
\begin{equation}
S^{(2)}_{\text{ghost}}\left[\varphi,{\bar c},c\right]=\sum^\infty_{s=2}(s-1)!\int_{\text{AdS}_{d+1}}\,
\bar{c}_{s-1}(x,\pl_{u})\,[\Box+\Lambda\,u\cdot\pl_u(u\cdot\pl_u+d-1)]\,c_{s-1}(x,u)\Big|_{u=0},
\end{equation}
where $c_{s-1}$, ${\bar c}_{s-1}$ are the ghosts associated to the Fronsdal field $\varphi_s$. Likewise, the cubic action is fixed by the first order deformation of the linearised gauge transformations, and reads
\begin{align}
S^{(3)}_{\text{ghost}}\left[\varphi,{\bar c},c\right]&=-\sum^\infty_{s_2=2}\frac{1}{s_2}\int_{\text{AdS}_{d+1}}\, \bar{c}_{s_2-1}(x,\pl_{u_2}) \left(\partial_{u_2}\cdot \nabla_2\right){\hat T}_{13}\left(c_{s_1-1},\varphi_{s_3}\right)\Big|_{u=0}\\&-\sum^\infty_{s_3=2}\frac{1}{s_3}\int_{\text{AdS}_{d+1}}\, \bar{c}_{s_3-1}(x,\pl_{u_3}) \left(\partial_{u_3}\cdot \nabla_3\right){\hat T}_{12}\left(c_{s_1-1},\varphi_{s_2}\right)\Big|_{u=0},\nonumber
\end{align}
where
\begin{align}
{\hat T}_{13}\left(\xi_1,\varphi_{s_3}\right)=&-\frac12\pl_{{\cal Y}_1}f^{TT}_{s_1,s_2,s_3}\,\xi_1\varphi_{s_3}-\frac14(d-2+2{\cal Y}_1\pl_{{\cal Y}_1})\pl_{{\cal Y}_1}\pl_{{\cal Y}_3}^2f^{TT}_{s_1,s_2,s_3} \xi_1^{(1)}\varphi_{s_3}^{\prime}\\ \nonumber & -\frac14(d-2+2{\cal Y}_3\pl_{{\cal Y}_3})\pl_{{\cal Y}_1}\pl_{{\cal Y}_2}^2\pl_{{\cal Y}_3}f^{TT}_{s_1,s_2,s_3}\partial_{U_2}\cdot \partial_{U_2}\xi_1\,U_3\cdot X_3\,\varphi_{s_3}^{\prime(2)}\\  \nonumber &-\frac18(d+2{\cal Y}_3\pl_{{\cal Y}_3})(d-2+2{\cal Y}_1\pl_{{\cal Y}_1})\pl_{{\cal Y}_1}\pl_{{\cal Y}_2}^2\pl_{{\cal Y}_3}^2f^{TT}_{s_1,s_2,s_3}\,\partial_{U_2}\cdot \partial_{U_2} \xi_1^{(1)}\varphi_{s_3}^{\prime(1)},
\end{align}
and
\begin{align}\label{F3int}
{\hat T}_{12}\left(\xi_1,\varphi_{s_2}\right)=&+\frac12\pl_{{\cal Y}_1}f^{TT}_{s_1,s_2,s_3}\,\xi_1\varphi_{s_2} +\frac14(d-2+2{\cal Y}_1\pl_{{\cal Y}_1})\pl_{{\cal Y}_1}\pl_{{\cal Y}_3}^2f^{TT}_{s_1,s_2,s_3}\partial_{U_3}\cdot \partial_{U_3} \xi_1^{(1)}\varphi_{s_2} \\ \nonumber
& -\frac14(d-2+2{\cal Y}_1\pl_{{\cal Y}_1})\pl_{{\cal Y}_1}\pl_{{\cal Y}_2}\pl_{{\cal Y}_3}^2f^{TT}_{s_1,s_2,s_3}\partial_{U_3}\cdot \partial_{U_3}\,\xi_1^{(1)} \,U_2\cdot X_2\,\varphi_{s_2}^{\prime(1)}\\ \nonumber
&-\frac18(d+2{\cal Y}_3\pl_{{\cal Y}_3})(d-2+2{\cal Y}_1\pl_{{\cal Y}_1})\pl_{{\cal Y}_1}\pl_{{\cal Y}_2}^2\pl_{{\cal Y}_3}^2f^{TT}_{s_1,s_2,s_3}\partial_{U_3}\cdot \partial_{U_3}\,\xi_1^{(1)} \varphi_{s_2}^{\prime(1)}\, \\ \nonumber
&-\frac12\pl_{{\cal Y}_1}\pl_{{\cal Y}_2}f^{TT}_{s_1,s_2,s_3}\,\xi_1\,U_2\cdot X_2\,\varphi_{s_2}^{\prime(1)},
\end{align}
which come from the first order deformation \eqref{1odef} of the linearsied gauge transformation induced by the de Donder gauge cubic couplings \eqref{osactionint}.

Having gauge fixed the action, in \S \tcb{\ref{subsec::bubuprop}} we also determine the complete form of the bulk-to bulk propagators for Fronsdal fields and ghosts in the de-Donder gauge, completing the results of \cite{Bekaert:2014cea} to include the gauge terms required for an off-shell source.

The bulk-to-bulk propagator for a spin-$s$ Fronsdal field in de Donder gauge is given by
\begin{align}
\Pi_{s}\left(x_1,u_1;x_2,u_2\right)&=\pi_{{\tilde \varphi}_1{\tilde \varphi}_2}\left(x_1,u_1;x_2,u_2\right)\\  \nonumber &+\frac{u^2_1}{2\left(d-3+2s\right)}\pi_{{\tilde \varphi}_1 \varphi^\prime_2}\left(x_1,u_1;x_2,u_2\right)
+\frac{u^2_2}{2\left(d-3+2s\right)}\pi_{\varphi^\prime_1{\tilde \varphi}_2}\left(x_1,u_1;x_2,u_2\right)\\ \nonumber
&+\frac{u^2_1u^2_2}{4\left(d-3+2s\right)^2}\pi_{\varphi^\prime_1 \varphi^\prime_2}\left(x_1,u_1;x_2,u_2\right),
\end{align}
where we decompose in traces. The traceless components are given in spectral form by
{\small \begin{subequations}
\begin{align}\label{eq:ManifestTrace2}
\pi_{{\tilde \varphi}_1{\tilde \varphi}_2}(x_1,w_1;x_2,w_2) &= \sum^{s}_{l=0} \int^{\infty}_{-\infty}  d\nu \; g^{{\tilde \varphi}_1{\tilde \varphi}_2}_{s,s-l}\left(\nu\right) \left(w_1\cdot\nabla\right)^{l} \left(w_2\cdot\nabla\right)^{l} \Omega_{\nu,s-l}(x_1,w_1;x_2,w_2),\\
\pi_{{\tilde \varphi}_1\varphi^\prime_2}(x_1,w_1;x_2,w_2) &= \sum^{s-2}_{l=0} \int^{\infty}_{-\infty}  d\nu \; g^{{\tilde \varphi}_1\varphi^\prime_2}_{s,s-l-2}\left(\nu\right) \left(w_1\cdot\nabla\right)^{l+2} \left(w_2\cdot\nabla\right)^{l} \Omega_{\nu,s-2-l}(x_1,w_1;x_2,w_2),\\
\pi_{\varphi^\prime_1{\tilde \varphi}_2}(x_1,w_1;x_2,w_2) &= \sum^{s-2}_{l=0} \int^{\infty}_{-\infty}  d\nu \; g^{\varphi^\prime_1{\tilde \varphi}_2}_{s,s-l-2}\left(\nu\right) \left(w_1\cdot\nabla\right)^{l} \left(w_2\cdot\nabla\right)^{l+2} \Omega_{\nu,s-2-l}(x_1,w_1;x_2,w_2),\\
\pi_{\varphi^\prime_1\varphi^\prime_2}(x_1,w_1;x_2,w_2) &= \sum^{s-2}_{l=0} \int^{\infty}_{-\infty}  d\nu \; g^{\varphi^\prime_1\varphi^\prime_2}_{s,s-l-2}\left(\nu\right) \left(w_1\cdot\nabla\right)^{l} \left(w_2\cdot\nabla\right)^{l} \Omega_{\nu,s-2-l}(x_1,w_1;x_2,w_2),
\end{align}
\end{subequations}} 
in terms of harmonic functions $\Omega$ (see \S \tcb{\ref{app::adsharm}}), whose coefficients read
\begin{subequations}
\begin{align}
g^{{\tilde \varphi}_1{\tilde \varphi}_2}_{s,s-l}\left(\nu\right)&=
\frac{64s(s-1)(d+2 s-5)}{l (-d+l-2 s+2)\left((d+2 s-2)^2+4 \nu ^2\right)} \\ \nonumber &  \hspace*{1.5cm} \times
\frac{c^{(s-2)}_{l-2}\left(\nu\right)}{ \left(d^2+4 d (l+s-2)+4 \left(-l^2+2 l (s-1)+\nu ^2+(s-2)^2\right)\right)^2}\\
g^{{\tilde \varphi}_1\varphi^\prime_2}_{s,s-2-l}\left(\nu\right)&=\frac{(l+2) (-d+l-2 s+4) \left((d+2 s-2)^2+4 \nu ^2\right) }{2 (d+2 s-5)}\,g^{{\tilde \varphi}_1{\tilde \varphi}_2}_{s,s-2-l}\left(\nu\right)\,,\\
g^{\varphi^\prime_1{\tilde \varphi}_2}_{s,s-2-l}\left(\nu\right)&=\frac{(l+2) (-d+l-2 s+4) \left((d+2 s-2)^2+4 \nu ^2\right) }{2 (d+2 s-5)}\,g^{{\tilde \varphi}_1{\tilde \varphi}_2}_{s,s-2-l}\left(\nu\right)\,,\\
g^{\varphi^\prime_1\varphi^\prime_2}_{s,s-2-l}\left(\nu\right)&=\frac{(l+2)^2 (-d+l-2 s+4)^2 \left((d+2 s-2)^2+4 \nu ^2\right)^2}{4 (d+2 s-5)^2}g^{{\tilde \varphi}_1{\tilde \varphi}_2}_{s,s-2-l}\left(\nu\right)\,.
\end{align}
\end{subequations}
The function $c^{(s-2)}_{l-2}\left(\nu\right)$ is given by \eqref{csl}.

The ghost bulk-to-bulk propagator is similarly given by:
{\small 
\begin{equation}
\Pi^{\text{gh.}}_{s-1}(x_1,w_1;x_2,w_2) = \sum^{s-1}_{l=0} \int^\infty_{-\infty}d\nu\,h_{s-1,s-1-l}\left(\nu\right) \left(w_1\cdot\nabla\right)^{l} \left(w_2\cdot\nabla\right)^{l} \Omega_{\nu,s-1-l}(x_1,w_1;x_2,w_2). 
\end{equation}}
with
\begin{equation}
h_{s-1,s-1-l}\left(\nu\right)=-c^{(s-1)}_{l}\left(\nu\right)\frac{d+2s-5}{\left(l-1\right)\left(2s+d-l-5\right)} \frac{1}{\nu^2+\left(s-3+\frac{d}{2}\right)^2}.
\end{equation}

We also determine the ghost bulk-to-boundary propagators in \S \tcb{\ref{subsubsec:ghbobu}}, which in the ambient space formalism read
\begin{equation}\label{ghostBtoB}
K^{\text{gh.}}_{s-1}(X,U;P,Z)=C_{d+s-1,s-1}X^2\frac{\left[\left(U \cdot Z\right)\left(-2P \cdot X\right)+2\left(U \cdot P\right)\left(Z \cdot X\right)\right]^{s-1}}{\left(-2 X \cdot P\right)^{d+2s-2}},
\end{equation}
and normalisation
\begin{equation}
C_{d+s-1,s-1} = \frac{\left(d+2s-3\right)\Gamma\left(d+s-2\right)}{2 \pi^{d/2}\Gamma\left(s+\tfrac{d}{2}\right)}\,.
\end{equation}
These are to accompany the bulk-to-boundary propagators for the associated spin-$s$ gauge fields, which were determined by Mikhailov in \cite{Mikhailov:2002bp}.

It is interesting to notice the factor of $X^2$ appearing in the bulk-to-boundary propagator for the ghosts. Although this factor does not affect the intrinsic projection, this is the correct ambient representation compatible with the fact that in our ambient space conventions ghost do not uplift directly as ambient space harmonic functions. This is due to the leftover mass term present in the ambient equations \eqref{gpambcon}. At the practical level one can reabsorb the additional factor of $X^2$ of \eqref{ghostBtoB} in the vertices redefining the fields in terms of harmonic representatives allowing a more uniform notation. However, since this issue is not important for the discussion of the present paper we do not discuss it further here.

\section{Notation, conventions and ambient space}
\label{sec::notconvamb}

In this work we consider higher-spin gauge theories in Euclidean anti-de Sitter (AdS$_{d+1}$) space, where the boundary dimension $d$ is general. 

Throughout we employ an operator formalism to manage the tensor indices (for a review see e.g. \cite{Sleight:2017krf}, whose conventions we adopt throughout), where fields are represented by generating functions. For example, a totally symmetric rank-$s$ bulk field $\varphi_{\mu_1...\mu_s}$ is represented as
\begin{equation}\label{genranks}
    \varphi_{\mu_1...\mu_s}\left(x\right)\:\rightarrow\: \varphi_s\left(x;u\right)=\frac{1}{s!} \varphi_{\mu_1...\mu_s}\left(x\right)u^{\mu_1}...u^{\mu_s},
\end{equation}
where we introduced the constant $\left(d+1\right)$-dimensional auxiliary vector $u^{\mu}$. 
In packaging totally symmetric tensor in generating functions as above, the action of the covariant derivatives is defined as a differential operator in both $x^\mu$ and $u^\mu$:
\begin{equation}\label{modincovder}
    \nabla_{\mu}\;\rightarrow\;\nabla_{\mu}-\frac{1}{2}\omega^{ab}_\mu L_a{}^b \; = \;\nabla_{\mu}+\omega^{ab}_\mu u_a \frac{\partial}{\partial u^b},
\end{equation}
where $u^a=e^a_{\mu}\left(x\right)u^\mu$ with viel-bein $e^a_{\mu}\left(x\right)$, $\omega^{ab}_\mu$ is the spin connection and $L_a{}^b$ the Lorentz tensors which are given by 
\begin{equation}
L_a{}^b=u^a\frac{\partial}{\partial u^b}-u^b\frac{\partial}{\partial u^a}, \qquad \partial_{u^a}u^b=\delta^{b}_a.
\end{equation}
In the following we shall work with contracted auxiliary variables $u^\mu= e^{\mu}_a\left(x\right)u^a$ and the associated derivative $\partial_{u^\mu}= e^a_{\mu}\left(x\right)\partial_{u^a}$. As a consequence of the vielbein postulate:
\begin{equation}
\left[\nabla_\mu,u^\nu\right]=0, \qquad \left[\partial_{u^\mu},\nabla_\nu\right]=0.
\end{equation}

The operator formalism is useful since it allows to translate tensor operations in terms of an operator calculus, which simplifies manipulations. For instance, the contraction:
\begin{equation}
    \varphi_{\mu_1...\mu_s}\left(x\right)\varphi^{\mu_1...\mu_s}\left(x\right)=s!\,\varphi_s\left(x;\partial_u\right)\varphi\left(x;u\right),
\end{equation}
and the operations:  divergence, symmetrised gradient, box, symmetrised metric, trace and spin are represented by the following operators:
\begin{align}
\textrm{divergence:  } 		& \nabla\cdot\partial_u,&
\textrm{sym.~gradient:  } 	& u\cdot\nabla,	&
	\textrm{box:  } 			& \Box , \\ 
	\textrm{sym.~metric:  } 	& u^2, \nonumber &
	\textrm{trace:  } 			& \partial_u^2,& 
	\textrm{spin:  } 			& u\cdot\partial_u .
\end{align}
In \S \tcb{\ref{app::opalg1}} we give the operator algebra. 

In this formalism, the usual Fierz-Pauli conditions for a symmetric bosonic spin-$s$  field of mass $m^2_sR^2=\Delta\left(\Delta-d\right)-s$ on AdS$_{d+1}$ are given by:
\begin{subequations}\label{fierzgen}
\begin{align}
\left(\Box-m^2_s\right)\varphi_s\left(x,u\right)\; & = \; 0,\\
\left(\partial_u \cdot \nabla \right)\varphi_s\left(x,u\right)\; & = \; 0,\\
\left(\partial_u \cdot \partial_u \right)\varphi_s\left(x,u\right)\; & = \; 0,
\end{align}
\end{subequations}

For $\Delta=s+d-2$, $\varphi_s$ is a gauge field and the system \eqref{fierzgen} is invariant under the gauge transformation
\begin{equation}\label{gt0}
\delta^{(0)}_{\xi_{s-1}}\varphi_{s}\left(x,u\right)=\left(u \cdot \nabla\right) \xi_{s-1}\left(x,u\right),
\end{equation}
with symmetric traceless rank-$\left(s-1\right)$ gauge parameter
\begin{equation}
\xi_{s-1}\left(x,u\right)=\frac{1}{\left(s-1\right)!}\xi_{\mu_1...\mu_{s-1}}u^{\mu_1}...u^{s-1}, \qquad \left(\partial_u \cdot \partial_u \right)\xi_{s-1}\left(x,u\right)\; = \; 0,
\end{equation}
which is \emph{on-shell}:
\begin{subequations}\label{onshgpcon}
\begin{align}
\left(\Box-m^2_\xi\right)\xi_{s-1}\left(x,u\right)\; & = \; 0,\\
\left(\partial_u \cdot \nabla \right)\xi_{s-1}\left(x,u\right)\; & = \; 0,\\
\left(\partial_u \cdot \partial_u \right)\xi_{s-1}\left(x,u\right)\; & = \; 0,
\end{align}
\end{subequations}
where $m^2_\xi R^2=\left(s-1\right)\left(s+d-2\right)$.

Fields which are symmetric and also traceless may furthermore be encoded in generating functions \eqref{genranks} of a null auxiliary vector $w^2=0$. In the operator calculus (see e.g. \cite{Joung:2013doa,Nutma:2014pua} and references therein) one replaces the partial derivative $\partial_w$ with the Thomas derivative \cite{10.2307/84634}:
\begin{equation}\label{thomasdint}
{\hat \partial}_{w^\mu} = \partial_{w^\mu}-\frac{1}{d-1+2 w \cdot \partial_w} w_{\mu} \partial^2_w,
\end{equation}
which preserves $w^2=0$. In this case the operator calculus simplifies to just four operators: 
\begin{align}
\Box, \qquad w \cdot \nabla, \qquad \nabla \cdot {\hat \partial}_w, \qquad w \cdot {\hat \partial}_w.
\end{align}
In \S \tcb{\ref{app::opalg2}} we give the corresponding operator algebra. 
 
On the conformal boundary of AdS$_{d+1}$, operators of non-trivial spin  can likewise be encoded in generating function notation. A totally symmetric spin-$s$ operator ${\cal O}_{i_1...i_s}$ at the boundary point $y^i$, $i=1,...,d$ can be packaged as 
\begin{equation}
    {\cal O}_{i_1...i_s}\left(y\right)\:\rightarrow\: {\cal O}_{s}\left(y;z\right)={\cal O}_{i_1...i_s}\left(y\right)z^{i_1}...z^{i_s},
\end{equation}
with the null auxiliary vector $z^2=0$. In this case the Thomas derivative is:\footnote{This is sometimes referred to as the Todorov differential operator \cite{Dobrev:1975ru} in the CFT literature.}
\begin{equation}
{\hat \partial}_{z^i} = \partial_{z^i} - \frac{1}{d-2+2 z \cdot \partial_z} z_i \partial^2_z.
\end{equation}

\subsection{Review: ambient space formalism}
\label{subsec::revamb}

In the ambient space formalism \cite{Dirac:1936fq} one regards AdS$_{d+1}$ as a co-dimension one hyper-surface
\begin{equation}
X^2+R^2=0, \label{hyp}
\end{equation}
in a $\left(d+2\right)$-dimensional ambient flat space-time, which we parameterise by Cartesian co-ordinates $X^A$ where $A=0,1,...,d+1$ with metric $\eta = \left(-++...+\right)$ to describe Euclidean AdS.

The totally symmetric spin $s$ field $\varphi_{\mu_1...\mu_s}$ of mass $m^2_sR^2=\Delta\left(\Delta-d\right)-s$ on AdS$_{d+1}$ is represented uniquely in ambient space by the tensor $\varphi_{A_1...A_s}$,
\begin{equation}
\varphi_{\mu_1...\mu_s}\left(x\right)=\frac{\partial X^{A_1}}{\partial x^{\mu_1}}...\frac{\partial X^{A_s}}{\partial x^{\mu_s}}\varphi_{A_1...A_s}\left(X\left(x\right)\right),
\end{equation}
subject to the following constraints \cite{Fronsdal:1978vb}:
\begin{itemize}
    \item {\bf Tangentiality} to surfaces of constant $\rho=\sqrt{-X^2}$:
    \begin{equation}\label{tangent}
       \left(X \cdot \partial_U\right) \varphi_{s}\left(X,U\right)=0, \quad i=1,...,s.
    \end{equation}
    \item The {\bf homogeneity} condition:
    \begin{equation}\label{homo}
        \left(X \cdot \partial_X+\mu\right)\varphi_{s}\left(X,U\right)=0, \quad \text{i.e.} \quad \varphi_{s}\left(\lambda X,U\right) = \lambda^{-\mu}\varphi_{s}\left( X,U\right),
    \end{equation}
    where we are free to choose either $\mu=\Delta$ or $\mu=d-\Delta$. In this work we take $\mu=\Delta$.
\end{itemize}
These constraints make sure that the ambient uplift of fields that live on the AdS manifold is well-defined and 1:1. Like in the previous section, in the above we introduced a generating function to represent the ambient field $\varphi_{A_1...A_s}$:
\begin{equation}\label{ambsygenfunc}
    \varphi_{A_1..A_s}\left(X\right)\:\rightarrow \: \varphi_s\left(X,U\right)=\frac{1}{s!} \varphi_{A_1..A_s}\left(X\right)U^{A_1}...U^{A_s},
\end{equation}
with constant ambient auxiliary vector $U^A$.

The above discussion extends to differential operators.  In the operator formalism, the ambient space representative $\nabla_A$ of the Levi-Civita connection $\nabla_\mu$ on AdS$_{d+1}$ is given by \cite{Metsaev:1995re,Bekaert:2010hk,Taronna:2012gb}:
\begin{equation}
\nabla_{A}={\cal P}_{A}^{B} \frac{\partial}{\partial X^B}-
\frac{X^B}{X^2}\Sigma_{AB}, \qquad X \cdot \nabla=0,
\end{equation}
with projector
\begin{equation}\label{ambproj}
{\cal P}^B_{A}=\delta^{B}_A-\frac{X_AX^B}{X^2}, \qquad \left(X \cdot {\cal P}\right)^B= 0, \qquad \left({\cal P} \cdot X\right)_A=0,
\end{equation}
and
\begin{equation}
    \Sigma_{AB}=U_A \frac{\partial}{\partial U^B}-U_B \frac{\partial}{\partial U^A}.
\end{equation}

In this framework, the intrinsic Fierz-Pauli system \eqref{fierzgen} is described by
\begin{subequations}\label{fpsysamb}
\begin{align}
\partial^2_X \varphi_s\left(X,U\right)&=0,\\
\left(\partial_X \cdot \partial_U\right)\varphi_s\left(X,U\right)&=0,\\
\left(\partial_U \cdot \partial_U\right)\varphi_s\left(X,U\right)&=0,
\end{align}
\end{subequations}
supplemented with the tangentiality and homogeneity conditions \eqref{tangent} and \eqref{homo}. 

For a spin $s$ gauge field $\Delta=s+d-2$, the gauge transformation \eqref{gt0} reads
\begin{equation}\label{ambgt}
\delta^{(0)}_{\xi_{s-1}}\varphi_s\left(X,U\right)=\left[U \cdot \partial_X+\frac{U \cdot X}{X^2}\left(U \cdot \partial_U - X \cdot \partial_X\right)\right]\xi_{s-1}\left(X,U\right),
\end{equation}
where the ambient representative of the gauge parameter is subject to the tangentiality and homogeneity conditions 
\begin{equation}\label{ttgpamb}
 \left(X \cdot \partial_U\right)\xi_{s-1}\left(X,U\right)=0, \qquad \left(X \cdot \partial_X+\mu-1\right)\xi_{s-1}\left(X,U\right)=0,
\end{equation}
and the on-shell constraints \eqref{onshgpcon} are represented as:
\begin{subequations}\label{gpambcon}
\begin{align}
\left[\partial^2_X+\frac{2}{X^2}\left(d-2+2U \cdot \partial_U\right)\right]\xi_{s-1}\left(X,U\right)&=0,\\
\left(\partial_X \cdot \partial_U\right)\xi_{s-1}\left(X,U\right)&=0,\\
\left(\partial_U \cdot \partial_U\right)\xi_{s-1}\left(X,U\right)&=0.
\end{align}
\end{subequations}
It is straightforward to verify that \eqref{ambgt} under the constraints \eqref{ttgpamb} and \eqref{gpambcon} leaves the Fierz system \eqref{fpsysamb} invariant.

Traceless fields living on the AdS manifold are represented by traceless ambient representatives with respect to the ambient metric $\eta_{AB}$, which themselves can be encoded in generating functions \eqref{ambsygenfunc} with null ambient auxiliary vector $W^2=0$. The Thomas derivative \eqref{thomasdint} reads
\begin{equation}\label{thomasdamb}
{\hat \partial}_{W^A} = \partial_{W^A}-\frac{1}{d-1+2 W \cdot \partial_W} W_{A}\partial^2_{W}.
\end{equation}
It is also sometimes useful to impose the constraints:
\begin{equation}\label{trauxamb}
X \cdot U = 0, \qquad X \cdot W=0,
\end{equation}
which take care of the tangentiality condition \eqref{tangent}. Preserving the above constraints in the operator calculus (which is given in \S \tcb{\ref{app::opalg3}} and \S \tcb{\ref{app::opalg4}}) requires the following modifications of $\partial_U$ and the Thomas derivative \eqref{thomasdamb}:
\begin{subequations}
\begin{align}
D_{U^A} &= \left({\cal P} \cdot \partial_U\right)_{A}\\
{\hat D}_{W^A} &=  \partial_{W^A}-\frac{1}{d-1+2 W \cdot {\cal P} \cdot  \partial_W} W_{A}\left(\partial_{W} \cdot {\cal P} \cdot \partial_{W}\right),
\end{align}
\end{subequations}
with projector \eqref{ambproj}.

\subsubsection*{The AdS boundary}

The ambient formalism can also be extended to the AdS boundary \cite{Dirac:1935zz,Dirac:1936fq,Fronsdal:1978vb,Bars:1998ph,Bekaert:2009fg,Costa:2011mg,Bekaert:2012vt}. As the boundary is approached, the hyperboloid \eqref{hyp} asymptotes to the light-cone. This limit does not yield a well-defined boundary metric, but one can obtain a finite limit by considering a projective cone of light-rays:
\begin{equation}
    P^A\equiv \epsilon X^A, \quad \epsilon \rightarrow 0.
\end{equation}
Because $X^2$ is fixed, these null co-ordinates satisfy:
\begin{equation}
    P^2=0, \qquad P\cong \lambda P, \qquad \lambda \neq 0,
\end{equation}
and are identified with the AdS boundary. For Euclidean AdS in Poincar\'e co-ordinates $x^\mu=\left(z,y^i\right)$, we have:
\begin{subequations}\label{poinamb}
\begin{align}
    X^0\left(x\right) &= R\frac{z^2+y^2+1}{2z} \\
    X^{d+1}\left(x\right) &=R\frac{1-z^2-y^2}{2z} \\
    X^i\left(x\right) & = \frac{Ry^i}{z}, 
\end{align}
\end{subequations}
and the boundary points are parameterised by the Poincar\'e section:
\begin{equation}\label{sect}
  P^{0}\left(y\right)=\frac{1}{2}\left(1+y^2\right), \quad P^{d+1}\left(y\right)=\frac{1}{2}\left(1-y^2\right), \quad P^i\left(y\right) = y^i.
  \end{equation}

A symmetric spin-$s$ boundary field $f_{i_1...i_s}\left(y\right)$ of scaling dimension $\Delta$ is assigned an ambient representative $f_{A_1 ...A_s}\left(P\right)$, which is traceless with respect to the ambient metric\footnote{This follows from the tracelessness of $f_{i_1...i_s}$.}
\begin{equation}
    \eta^{AB}f_{A_1 ...A_s}=0 \label{tlessbound}
\end{equation}
and scales as
\begin{equation}
    f_{A_1 ...A_s}\left(\lambda P\right)=\lambda^{-\Delta}f_{A_1 ...A_s}\left(P\right), \qquad \lambda > 0.
\end{equation}
Like for the ambient description of bulk fields detailed above, we require that $f_{A_1 ...A_s}$ is tangent to the light-cone:
\begin{equation}
    P^{A_1}f_{A_1 ...A_s}\left(P\right)=0.\label{transbound}
\end{equation}
However, because $P^2=0$, there is an additional redundancy
\begin{subequations}\label{extred}
\begin{align}
 & \hspace*{3.5cm} f_{A_1...A_s}(P) \rightarrow f_{A_1...A_s}(P) + P_{\left(A_1\right.}\Lambda_{\left. A_2 ... A_s \right)},\\
 & P^{A_1}\Lambda_{ A_1 ... A_{s-1}} = 0, \quad \Lambda_{ A_1 ... A_{s-1}}(\lambda P) = \lambda^{-(\Delta+1)}\Lambda_{ A_1 ... A_{s-1}}(P), \quad \eta^{A_1A_2}\Lambda_{A_1 ... A_{s-1}} = 0,
\end{align}
\end{subequations}
which, combined with \eqref{transbound}, eliminates the extra two degrees of freedom per index of $f_{A_1...A_s}$.

The operator formalism also extends to ambient boundary fields, where we have:
\begin{equation}
    f_{A_1...A_s}\left(P\right)\:\rightarrow\: f_s\left(P,Z\right)=\frac{1}{s!}f_{A_1...A_s}\left(P\right)Z^{A_1}...Z^{A_s}, \quad Z^2=0, \quad P\cdot Z=0,
\end{equation}
with the null ambient auxiliary vector $Z^2=0$ imposing the traceless condition \eqref{tlessbound} and it is useful to introduce the constraint $P\cdot Z=0$ which implements tangentiality to the light-cone \eqref{transbound}.

\subsection{Functionals in de Donder gauge}

In this paper we are interested in providing a convenient formalism to deal with AdS Feynman rules for higher-spin gauge fields in the de-Donder gauge, including ghosts. It is thus a key step to describe in detail the corresponding tensor calculus.

An off-shell Langrangian for a spin-$s$ gauge field $\varphi_s$ freely propagating on AdS$_{d+1}$ is given by the Fronsdal action \cite{Fronsdal:1978rb}, which in the operator formalism reads
\begin{equation}\label{Fronsdalaction}
S^{(2)}\left[\varphi_s\right] = \frac{s!}{2}\int_{\text{AdS}_{d+1}} \varphi_{s}\left(x,\partial_u\right)\left(1-\frac{1}{4}u^2 \partial_u \cdot \partial_u\right){\hat {\cal F}}_s\,\varphi_{s}\left(x,u\right)\Big|_{u=0}
\end{equation}
with Fronsdal operator
\begin{subequations}\label{Fronsdaltensor}
\begin{align} 
{\hat {\cal F}}_{s}
& =
\Box- m^2_s-u^2(\partial_u\cdot \partial_u)
-\;(u\cdot \nabla){\hat {\cal D}},\\
{\hat {\cal D}} & = (\nabla\cdot\partial_u)-\frac{1}{2}(u\cdot \nabla)
(\partial_u\cdot \partial_u),
\end{align}
\end{subequations}
where $m^2_sR^2=\left(s-2\right)\left(s+d-2\right)-s$ and ${\hat {\cal D}}$ is the de Donder operator. Fronsdal fields have vanishing double trace:
\begin{equation}
\left(\partial_u \cdot \partial_u\right)^2\varphi_{s}\left(x,u\right)=0,
\end{equation}
and gauge transformation:
\begin{equation}\label{gaugetransf}
\delta^{(0)}_{\xi}\varphi_s\left(x,u\right)=u \cdot \nabla \xi_{s-1}\left(x,u\right),
\end{equation}
with symmetric and traceless rank $s-1$ gauge parameter: $\left(\partial_u \cdot \partial_u\right) \xi_{s-1}\left(x,u\right)=0$. The equation of motion derived from the free action \eqref{Fronsdalaction} is given by 
\begin{equation}
\left(1-\frac{1}{4}u^2 \partial_u \cdot \partial_u\right){\cal F}_s\left(x,u, \nabla, \partial_u\right)\varphi_{s}\left(x,u\right)= 0.
\end{equation}
Since the operator $\left(1-\frac{1}{4}u^2 \partial_u \cdot \partial_u\right)$ is invertible,\footnote{Its inverse on doubly traceless fields can be easily computed, and is given by $$\left(1-\frac{1}{4}u^2 \partial_u \cdot \partial_u\right)^{-1}=\frac12\frac{u^2\pl_{u}^2}{d-5+2u\cdot\pl_u}$$.} this is equivalent to 
\begin{equation}\label{eomfronsdal}
{\cal F}_s\left(x,u, \nabla, \partial_u\right)\varphi_{s}\left(x,u\right)= 0.
\end{equation}
In this work we consider Fronsdal fields in the de Donder gauge:
\begin{equation}\label{dedonder}
{\hat {\cal D}} \varphi_s=0,
\end{equation}
where the equation of motion \eqref{eomfronsdal} takes the form
\begin{equation}\label{dedondeom}
\left(\Box- m^2_s-u^2\partial_u\cdot \partial_u\right)\varphi_s\left(x,u\right)\approx 0\,,
\end{equation}
and can be inverted off-shell.
The residual gauge freedom is given by \eqref{gaugetransf} with on-shell gauge parameters:
\begin{align}\label{dedondergaugetran}
\left(\Box-m^2_{\xi}\right)\xi_{s-1}\left(x,u\right)=0,
\end{align}
where $m^2_{\xi}R^2=\left(s-1\right)\left(s+d-2\right)$. Any further gauge fixing would thus be an on-shell gauge fixing. For example, the remaining freedom \eqref{dedondergaugetran} can be used to eliminate the trace of the de Donder field \eqref{dedonder} on-shell, such that it becomes transverse and traceless ($TT$):
\begin{equation}\label{ttintconds}
\left(\nabla \cdot \partial_u \right)\varphi_s\left(x,u\right)=0, \qquad \left(\partial_u \cdot \partial_u \right)\varphi_s\left(x,u\right)=0.
\end{equation}
This recovers the Fierz Pauli system \eqref{fierzgen}.

It is sometimes useful to express the double-traceless Fronsdal field in terms of its irreducible components 
\begin{equation}\label{irredecomp}
\varphi_s\left(x,u\right) = {\tilde \varphi}_{s}\left(x,u\right)+\frac{u^2}{2\left(d-3+2s\right)}  \varphi^{\prime}_s\left(x,u\right)
\end{equation}
where
\begin{equation}
\left(\partial_u \cdot \partial_u\right) \varphi_s\left(x,u\right) = \varphi^{\prime}_s\left(x,u\right), \qquad \left(\partial_u \cdot \partial_u\right){\tilde \varphi}_{s}\left(x,u\right)=\left(\partial_u \cdot \partial_u\right)\varphi^{\prime}_s\left(x,u\right)=0.
\end{equation}
In the de Donder gauge, the two traceless fields $\varphi^{\prime}_s$ and ${\tilde \varphi}_{s}$ completely decouple, with only the de Donder gauge condition \eqref{dedonder} relating them. The equation of motion \eqref{dedondeom} decomposes as:
\begin{subequations}
\begin{align}
\left(\Box-m_s^2\right){\tilde \varphi}_s(x,u)&=0\\
\left(\Box-m_t^2\right)\varphi^\prime_s(x,u)&=0
\end{align}
\end{subequations}
where $m_t^2R^2=s^2+(d-1)s-2$.

\subsubsection*{de Donder functionals in ambient space}

 The double traceless of the Fronsdal field also extends to its ambient representative:
 \begin{equation}
 \left(\partial_U \cdot \partial_U \right)^2\varphi_s\left(X,U\right)=0,
 \end{equation}
which supplements the tangentiality and homogeneity conditions \eqref{tangent} and \eqref{homo}.

Using the identities \eqref{actcodamrep} for the action of the ambient representative of the covariant derivative on tangent fields, it is straightforward to write down the ambient counterpart of the de Donder gauge condition \eqref{dedonder} and equation of motion \eqref{dedondeom}:
\begin{subequations}\label{dedondeomamb}
\begin{align}
{\hat {\cal D}}\varphi_s(X,U)&=\left\{\pl_U\cdot\pl_X-\frac12\left[U\cdot\pl_X+\frac{U\cdot X}{X^2}(d-2+2\,U\cdot\pl_U)\right]\pl_U^2\right\}\varphi_s(X,U)=0 \\ 
{\hat {\cal F}}_s\varphi_s(X,U)&=\left[\partial^2_X+2\frac{U\cdot X}{X^2}\,\pl_U\cdot\pl_{X}+\frac{U^2}{X^2}\pl_U^2\right]\varphi_s(X,U)=0,
\end{align}
\end{subequations}
with gauge transformation
\begin{align}
& \delta^{(0)}_{\xi}\varphi_s(X,U)=\left[U\cdot\pl_X+\frac{X\cdot U}{X^2}\,(d-2+2\,U\cdot\pl_U)\right]\xi_{s-1}(X,U),
\end{align}
where the ambient representative of the gauge parameter \eqref{dedondergaugetran} is subject to the on-shell constraints
\begin{subequations}
\begin{align}
 \left[\Box+2\,\frac{U\cdot X}{X^2}\,\pl_U\cdot\pl_X+\frac2{X^2}(d-2+2\,U\cdot\pl_U)\right]\xi_{s-1}(X,U)&=0,\\
 \left(\partial_U \cdot \partial_U\right)\xi_{s-1}(X,U)&=0,
\end{align}
\end{subequations}
in addition to the tangentiality and homogeneity conditions \eqref{ttgpamb}.

The decomposition \eqref{irredecomp} of the Fronsdal tensor into irreducible components takes the form
\begin{equation}
\varphi_s\left(X,U\right)={\tilde \varphi}_s\left(X,U\right)+\frac{U \cdot {\cal P} \cdot U}{2\left(d-3+2s\right)} \varphi^\prime_s\left(X,U\right),
\end{equation}
where
\begin{equation}
\left(\partial_U \cdot \partial_U \right)\varphi_s\left(X,U\right)=\varphi^\prime_s\left(X,U\right), \qquad \left(\partial_U \cdot \partial_U \right){\tilde \varphi}_s\left(X,U\right)=\left(\partial_U \cdot \partial_U \right)\varphi^\prime_s\left(X,U\right)=0.
\end{equation}
Both components are subject to the same homogeneity and tangentiality conditions \eqref{homo} and \eqref{tangent}.
In de Donder gauge, they are governed by the ambient equations of motions
\begin{subequations}
\begin{align}
\left[\partial^2_X+2\frac{U\cdot X}{X^2}\,\pl_U\cdot\pl_{X}\right] {\tilde \varphi}_s(X,U)&=0\\
\left[\Box+2\,\frac{U\cdot X}{X^2}\,\pl_U\cdot\pl_X+\frac2{X^2}(d+2\,U\cdot\pl_U)\right]\varphi^\prime_s(X,U)&=0.
\end{align}
\end{subequations}

\section{Off-shell cubic couplings in de Donder gauge}

\subsection{Fronsdal field cubic couplings}
\label{subsec::ffcc}

 The cubic action for traceless and transverse fields \eqref{ttintconds} in the type A higher-spin gauge theory on AdS$_{d+1}$ is given in the ambient space formalism by \cite{Sleight:2016dba,Sleight:2016xqq}\footnote{See \cite{Boulanger:2008tg,Bekaert:2010hk,Joung:2011ww,Joung:2013doa,Joung:2013nma} for previous studies and classifications of metric-like cubic vertices of totally symmetric higher-spin gauge fields in AdS. See also \cite{Francia:2016weg} for some recent developments on cubic couplings in the Maxwell-like formulation \cite{Francia:2011qa,Campoleoni:2012th} of higher-spin gauge fields.}
\begin{subequations}
\begin{align} \label{ttaction}
S^{(3)}_{TT} &= \sum_{s_1\geq s_2\geq s_3} {\cal V}^{TT}_{s_1,s_2,s_3}\\ \label{ttvertex}
{\cal V}^{TT}_{s_1,s_2,s_3}&=f^{TT}_{s_1,s_2,s_3} \varphi_{s_1}\left(X_1,U_1\right)\varphi_{s_2}\left(X_2,U_2\right)\varphi_{s_s}\left(X_3,U_3\right)\Big|_{X_i=X}, \\ \nonumber
f^{TT}_{s_1,s_2,s_3}&= g_{s_1,s_2,s_3}{\cal Y}^{s_1}_1{\cal Y}^{s_2}_2{\cal Y}^{s_3}_3
\end{align}
\end{subequations}
where 
\begin{align}\label{yis}
    \mathcal{Y}_1&=\pl_{U_1}\cdot\pl_{X_2}\,,&\mathcal{Y}_2&=\pl_{U_2}\cdot\pl_{X_3}\,,&\mathcal{Y}_3&=\pl_{U_3}\cdot\pl_{X_1},
\end{align}
and with coupling constant \cite{Sleight:2016dba,Sleight:2016xqq}
\begin{equation}\label{typeacc}
   g_{s_1,s_2,s_3}=\frac{1}{\sqrt{N}}\frac{\pi ^{\frac{d-3}{4}}2^{\tfrac{3 d-1+s_1+s_2+s_3}{2}}}{ \Gamma (d+s_1+s_2+s_3-3)}  \prod_{i=1}^3\sqrt{\frac{\Gamma(s_i+\tfrac{d-1}{2})}{\Gamma\left(s_i+1\right)}}.
\end{equation}
where $N$ is the only free parameter and is related via holography to the number of CFT degrees of freedom. The off-shell completion of the traceless and transverse cubic action \eqref{ttaction} can be determined using the Noether procedure, which at the cubic order requires:
\begin{equation}\label{cubcinoe}
\delta^{(1)}_{\xi}S^{(2)}+\delta^{(0)}_{\xi}S^{(3)}=0,
\end{equation}
where $S^{(2)}$ is the free off-shell type A action \eqref{typeafree} for Fronsdal fields with linearised gauge transformations \eqref{gaugetransf}. The $S^{(3)}$ is the off-shell cubic action we would like to determine and $\delta^{(1)}_{\xi}$ the corresponding first order deformation of the linearised gauge transformations.

Modulo the free Fronsdal equations of motion \eqref{eomfronsdal}, the condition \eqref{cubcinoe} reads:
\begin{equation}\label{d0cubic}
\delta^{(0)}_{\xi}S^{(3)} = 0,
\end{equation}
which we may solve in de Donder gauge for $S^{(3)}$ using the traceless and transverse action \eqref{ttaction} as a starting point.

Since the de Donder gauge condition \eqref{dedonder} allows to replace gradients with traces, $S^{(3)}$ can be decomposed in traces of the constituent fields:
\begin{subequations}\label{osaction}
\begin{align}
S^{(3)}&=\sum_{s_1\geq s_2 \geq s_3} {\cal V}_{s_1,s_2,s_3}\\ \label{offshell0}
{\cal V}_{s_1,s_2,s_3}&={\cal V}^{TT}_{s_1,s_2,s_3}+{\cal V}^{(1)}_{s_1,s_2,s_3}+{\cal V}^{(2)}_{s_1,s_2,s_3}+{\cal V}^{(3)}_{s_1,s_2,s_3},
\end{align}
\end{subequations}
where
\begin{subequations}
\begin{align}
{\cal V}^{(1)}_{s_1,s_2,s_3}&= f^{(1)}_{s_1}\varphi^\prime_{s_1}\varphi_{s_2}\varphi_{s_3}+\text{cyclic}\\
{\cal V}^{(2)}_{s_1,s_2,s_3}&= f^{(2)}_{s_1,s_2}\varphi^\prime_{s_1}\varphi^\prime_{s_2}\varphi_{s_3}+\text{cyclic}\\
{\cal V}^{(3)}_{s_1,s_2,s_3}&=f^{(3)}_{s_1,s_2,s_3}\varphi^\prime_{s_1}\varphi^\prime_{s_2}\varphi^\prime_{s_3},
\end{align}
\end{subequations}
for some functions $f^{(1)}_{s_i}$, $f^{(2)}_{s_i,s_j}$ and $f^{(3)}_{s_1,s_2,s_3}$ of the operators \eqref{yis}, which we determine in the following.  

For all three constituent fields in de Donder gauge, variation of the TT vertex \eqref{ttvertex} under a linearised spin-$s_1$ gauge transformation is given by 
\begin{align}
\delta^{(0)}_{\xi_{s_1-1}}\mathcal{V}^{TT}_{s_1,s_2,s_3}=&-\frac12\left(\pl_{X_2}^2-\pl_{X_3}^2\right)(\pl_{{\cal Y}_1}f^{TT}_{s_1,s_2,s_3})\,\xi^{(0)}_{s_1-1}\varphi_{s_2}\varphi_{s_3}\nonumber\\
& +(\pl_{{\cal Y}_1}f^{TT}_{s_1,s_2,s_3})\,U_1\cdot X_1\,\pl_{U_1}\cdot\pl_{X_1}\,\xi^{(1)}_{s_1-1}\varphi_{s_2}\varphi_{s_3} \nonumber
\\
&-(d-2+2{\cal Y}_1\pl_{{\cal Y}_1})(d-1+{\cal Y}_i\pl_{{\cal Y}_i})(\pl_{{\cal Y}_1}\pl_{{\cal Y}_3}^2f^{TT}_{s_1,s_2,s_3})\,\xi^{(1)}_{s_1-1}\varphi_{s_2}\varphi^\prime_{s_3}\nonumber\\ \label{tttranoff}
&-(d-2+2{\cal Y}_1\pl_{{\cal Y}_1})(\pl_{{\cal Y}_3}\pl_{{\cal Y}_1}f^{TT}_{s_1,s_2,s_3})\,\pl_{U_3}\cdot\pl_{X_3}\xi^{(1)}_{s_1-1}\varphi_{s_2}\varphi_{s_3}\,,
\end{align}
where for convenience we introduced the notation:
\begin{equation}\label{fnxu}
    f^{(n)}(X,U)\equiv\frac1{(X^2)^n}\,f(X,U)\,,
\end{equation}
for some field $f(X,U)$  in ambient space. In \eqref{app::opalg3} we give some useful identities. The first line of the variation \eqref{tttranoff} is the standard off-shell transformation for traceless and transverse fields \cite{Joung:2013nma,Sleight:2016xqq}, while the remaining terms denote the corrections to the latter in de Donder gauge.

The approach we take to determine the off-shell cubic couplings \eqref{offshell0} is to begin with traceless and transverse fields \eqref{ttintconds} and uplift them to de Donder gauge one-by-one. We express this schematically as follows:
\begin{equation}\label{schemsolve}
{\cal V}^{TT}_{s_1,s_2,s_3} \quad
\underset{\varphi_{s_1}}{\overset{\circled{\tiny 1}}{\bm\longrightarrow}}
 \quad {\cal V}^{(1)}_{s_1,s_2,s_3} \quad \underset{\varphi_{s_1},\,\varphi_{s_2}}{\overset{\circled{\tiny 2}}{\bm\longrightarrow}} \quad {\cal V}^{(2)}_{s_1,s_2,s_3} \quad \underset{\varphi_{s_1},\,\varphi_{s_2},\,\varphi_{s_3}}{\overset{\circled{\tiny 3}}{\bm\longrightarrow}} \quad {\cal V}^{(3)}_{s_1,s_2,s_3},
\end{equation}
where the notation {\small $\underset{\varphi_{s_i}}{\longrightarrow}$} signifies solving \eqref{d0cubic} at each step for the Fronsdal field $\varphi_{s_i}$ in de Donder gauge. In taking this approach to solve at a given order $n$ in traces for ${\cal V}^{(n)}_{s_1,s_2,s_3}$, the corrections ${\cal V}^{(m>n)}_{s_1,s_2,s_3}$ that are higher order in the traces of the fields do not contribute since only $n$ fields are uplifted to de Donder gauge and the rest are kept traceless and transverse. 

We thus first solve for ${\cal V}^{(1)}_{s_1,s_2,s_3}$ where $\varphi_{s_1}$ is in de Donder gauge with $\varphi_{s_2}$ and $\varphi_{s_3}$ traceless and transverse. In this case, the first line of the transformation \eqref{tttranoff} of the TT vertex is vanishing on the free equations of motion of $\varphi_{s_2}$ and $\varphi_{s_3}$, while the final two lines are vanishing owing to the traceless and transverse conditions \eqref{ttintconds}. The term on the second line is non-vanishing and must thus be cancelled by the variation of ${\cal V}^{(1)}_{s_1,s_2,s_3}$. Since the term to be cancelled is proportional to the divergence of the gauge parameter, the appropriate counter-term is simply given by: 
\begin{equation}
-\frac12\,(\pl_{Y_1}f^{TT}_{s_1,s_2,s_3})\,U_1\cdot X_1\,\varphi^{\prime(1)}_{s_1}\varphi_{s_2}\varphi_{s_3}\,.
\end{equation}
By cyclising, we thus have:
\begin{multline}
{\cal V}^{(1)}_{s_1,s_2,s_3}=-\frac12\,(\pl_{{\cal Y}_1}f^{TT}_{s_1,s_2,s_3})\,U_1\cdot X_1\,\varphi_{s_1}^{\prime(1)}\varphi_{s_2}\varphi_{s_3}-\frac12\,(\pl_{{\cal Y}_2}f^{TT}_{s_1,s_2,s_3})\varphi_{s_1}\,U_2\cdot X_2\,\varphi_{s_2}^{\prime(1)}\varphi_{s_3}\\-\frac12\,(\pl_{{\cal Y}_3}f^{TT}_{s_1,s_2,s_3})\varphi_{s_1}\varphi_{s_2}\,U_3\cdot X_3\,\varphi_{s_3}^{\prime(1)}\,.
\end{multline}

We can now proceed to $\circled{\tiny 2}$ and solve for ${\cal V}^{(2)}_{s_1,s_2,s_3}$ by compensating terms coming from the variation
\begin{align}
  \delta^{(0)}_{\xi_{s_1-1}}\left({\cal V}^{TT}_{s_1,s_2,s_3}+{\cal V}^{(1)}_{s_1,s_2,s_3}\right)=  &-\pl_{{\cal Y}_1}\pl_{{\cal Y}_2}f^{TT}_{s_1,s_2,s_3}\, U_1\cdot X_1\,\pl_{U_1}\cdot\pl_{X_1}\xi_{s_1-1}^{(1)}\,U_2\cdot X_2\,\varphi_{s_2}^{\prime(1)}\varphi_{s_3}\\\nonumber
    &-\frac12(d-2+2{\cal Y}_2\pl_{{\cal Y}_2})\pl_{{\cal Y}_1}^2\pl_{{\cal Y}_2}f^{TT}_{s_1,s_2,s_3}\,\pl_{U_1}\cdot\pl_{X_1}\,\xi_{s_1-1}\,U_2\cdot X_2\,\varphi_{s_2}^{\prime(2)}\varphi_{s_3}\,.
\end{align}
Like in the previous step, since each term is proportional to the divergence of the gauge parameter we can straightforwardly write down the counter-term
\begin{multline}
   \frac12\,\pl_{{\cal Y}_1}\pl_{{\cal Y}_2}f^{TT}_{s_1,s_2,s_3}\, U_1\cdot X_1\,\varphi^{\prime(1)}_{s_1}\,U_2\cdot X_2\,\varphi_{s_2}^{\prime(1)}\varphi_{s_3}\\
    +\frac14(d-2+2{\cal Y}_2\pl_{{\cal Y}_2})\pl_{{\cal Y}_1}^2\pl_{{\cal Y}_2}f^{TT}_{s_1,s_2,s_3}\,\varphi^\prime_{s_1}\,U_2\cdot X_2\,\varphi_{s_2}^{\prime(2)}\varphi_{s_3},
\end{multline}
which gives
\begin{multline}
    {\cal V}^{(2)}_{s_1,s_2,s_3}=\frac12\,\pl_{{\cal Y}_1}\pl_{{\cal Y}_2}f^{TT}_{s_1,s_2,s_3}\, U_1\cdot X_1\,\varphi^{\prime(1)}_{s_1}\,U_2\cdot X_2\,\varphi_{s_2}^{\prime(1)}\varphi_{s_3}+\text{cyclic}\\
    +\frac14(d-2+2{\cal Y}_2\pl_{{\cal Y}_2})\pl_{{\cal Y}_1}^2\pl_{{\cal Y}_2}f^{TT}_{s_1,s_2,s_3}\,\varphi^\prime_{s_1}\,U_2\cdot X_2\,\varphi_{s_2}^{\prime(2)}\varphi_{s_3}+\text{cyclic}\,.
\end{multline}
Proceeding to the final step $\circled{\tiny 3}$, we determine ${\cal V}^{(3)}_{s_1,s_2,s_3}$ by cancelling terms coming from the variation: 
\begin{align}\nonumber
&\delta^{(0)}_{\xi_{s_1-1}}\left({\cal V}^{TT}_{s_1,s_2,s_3} +{\cal V}^{(1)}_{s_1,s_2,s_3} + {\cal V}^{(2)}_{s_1,s_2,s_3}\right)\\  &\hspace*{2cm}=\frac12\,\pl_{X_1}\cdot\pl_{X_2}\,\pl_{{\cal Y}_1}\pl_{{\cal Y}_2}\pl_{{\cal Y}_3}f^{TT}_{s_1,s_2,s_3}\,\xi_{s_1-1}\,U_2\cdot X_2\,\varphi_{s_2}^{\prime(1)}\,U_3\cdot X_3\,\varphi_{s_3}^{\prime(1)}\\\nonumber
  & \hspace*{2cm}  +\frac14\,\pl_{X_1}\cdot\pl_{X_2}\,(d-2+2{\cal Y}_3\pl_{{\cal Y}_3})\pl_{{\cal Y}_1}\pl_{{\cal Y}_2}^2\pl_{{\cal Y}_3}f^{TT}_{s_1,s_2,s_3}\,\xi_{s_1-1}\varphi_{s_2}^\prime\,U_3\cdot X_3\varphi_{s_3}^{\prime(2)}\\\nonumber
 & \hspace*{2cm}   -\frac12(d-2+2{\cal Y}_1\pl_{{\cal Y}_1})\pl_{{\cal Y}_1}\pl_{{\cal Y}_2}\pl_{{\cal Y}_3}f^{TT}_{s_1,s_2,s_3}\,\xi_{s_1-1}^{(1)}\,U_2\cdot X_2\varphi_{s_2}^{\prime(1)}\,U_3\cdot X_3\varphi_{s_3}^{\prime(1)}\\\nonumber
& \hspace*{2cm}    +\frac12(d-1+{\cal Y}_i\pl_{{\cal Y}_i})(d-2+2{\cal Y}_1\pl_{{\cal Y}_1})\pl_{{\cal Y}_1}\pl_{{\cal Y}_2}\pl_{{\cal Y}_3}f^{TT}_{s_1,s_2,s_3}\,\xi_{s_1-1}\,U_2\cdot X_2\,\varphi_{s_2}^{\prime(1)}\,U_3\cdot X_3\,\varphi_{s_3}^{\prime(1)}\\\nonumber
 & \hspace*{2cm}  +\frac14(d+{\cal Y}_i\pl_{{\cal Y}_i})(d-2+2{\cal Y}_1\pl_{{\cal Y}_1})(d-2+2{\cal Y}_3\pl_{{\cal Y}_3})\pl_{{\cal Y}_1}\pl_{{\cal Y}_2}^2\pl_{{\cal Y}_3}f^{TT}_{s_1,s_2,s_3}\,\xi_{s_1-1}\varphi_{s_2}^\prime\,U_3\cdot X_3\,\varphi_{s_3}^{\prime(2)}\\\nonumber
& \hspace*{2cm}    -\frac14(d-2+2{\cal Y}_1\pl_{{\cal Y}_1})(d+2{\cal Y}_3\pl_{{\cal Y}_3})\pl_{{\cal Y}_1}\pl_{{\cal Y}_2}^2\pl_{{\cal Y}_3}^2f^{TT}_{s_1,s_2,s_3}\,\xi_{s_1-1}^{(1)}\varphi_{s_2}^\prime[U_3\cdot X_3\,\partial_{U_3} \cdot \partial_{X_3}-2]\varphi_{s_3}^{\prime(2)}\\\nonumber
& \hspace*{2cm}    -\frac12(d-2+2{\cal Y}_1\pl_{{\cal Y}_1})\pl_{{\cal Y}_1}\pl_{{\cal Y}_2}\pl_{{\cal Y}_3}^2f^{TT}_{s_1,s_2,s_3}\,\xi_{s_1-1}^{(1)}\,U_2\cdot X_2\,\varphi_{s_2}^{\prime(1)}\,U_3\cdot X_3\,\partial_{U_3} \cdot \partial_{X_3} \varphi_{s_3}^{\prime(1)}\\\nonumber
& \hspace*{2cm} -\frac12(\pl_{Y_1}\pl_{Y_2}f^{TT}_{s_1,s_2,s_3})\xi_{s_1-1}\,U_2\cdot X_2\,\varphi_{s_2}^{\prime(1)}\,\Box_3\varphi_{s_3}\\\nonumber
& \hspace*{2cm}+\frac12 \pl_{U_3}\cdot\pl_{X_3}(d-2+2Y_1\pl_{Y_1})(\pl_{Y_1}\pl_{Y_2}\pl_{Y_3}f^{TT}_{s_1,s_2,s_3})\xi_{s_1-1}^{(1)}\,U_2\cdot X_2\,\varphi_{s_2}^{\prime(1)}\varphi_{s_3}\\\nonumber
& \hspace*{2cm}+\frac12(d+Y_i\pl_{Y_i})(d-2+2Y_1\pl_{Y_1})(\pl_{Y_1}\pl_{Y_2}\pl_{Y_3}^2f^{TT}_{s_1,s_2,s_3})\xi_{s_1-1}^{(1)}\,U_2\cdot X_2\,\varphi_{s_2}^{\prime(1)}\varphi_{s_3}^\prime\\\nonumber
& \hspace*{2cm}+\frac12\,\pl_{U_2}\cdot\pl_{X_2}(d-2+2Y_3\pl_{Y_3})\pl_{Y_1}\pl_{Y_2}\pl_{Y_3}f^{TT}_{s_1,s_2,s_3}\, \xi_{s_1-1}\varphi_{s_2}\,U_3\cdot X_3\,\varphi_{s_3}^{\prime(2)}\\\nonumber
& \hspace*{2cm}+\frac12\,(d+Y_i\pl_{Y_i})(d-2+2Y_3\pl_{Y_3})\pl_{Y_3}\pl_{Y_1}\pl_{Y_2}^2f^{TT}_{s_1,s_2,s_3}\,\xi_{s_1-1}\varphi_{s_2}^\prime\,U_3\cdot X_3\,\varphi_{s_3}^{\prime(2)}\\\nonumber
& \hspace*{2cm}+\frac12 \pl_{U_2}\cdot\pl_{X_2}\,(d-2+2Y_3\pl_{Y_3})(d-2+2Y_1\pl_{Y_1})\pl_{Y_1}\pl_{Y_2}\pl_{Y_3}^2f^{TT}_{s_1,s_2,s_3}\,\xi_{s_1-1}^{(1)}\varphi_{s_2}\varphi_{s_3}^{\prime(1)}\\\nonumber
& \hspace*{2cm}+\frac12(d+1+Y_i\pl_{Y_i})(d-2+2Y_3\pl_{Y_3})(d-2+2Y_1\pl_{Y_1})\pl_{Y_1}\pl_{Y_2}^2\pl_{Y_3}^2f\,\xi_{s_1-1}^{(1)}\varphi_{s_2}^\prime\varphi_{s_3}^{\prime(1)}\\\nonumber
& \hspace*{2cm}+\pl_{U_2}\cdot\pl_{X_2}(d-2+2Y_1\pl_{Y_1})\pl_{Y_1}\pl_{Y_2}\pl_{Y_3}^2f^{TT}_{s_1,s_2,s_3}\xi_{s_1-1}^{(1)}\varphi_{s_2}\varphi_{s_3}^{\prime(1)}\\\nonumber
& \hspace*{2cm}+(d+1+Y_i\pl_{Y_i})(d-2+2Y_1\pl_{Y_1})\pl_{Y_1}\pl_{Y_2}^2\pl_{Y_3}^2f^{TT}_{s_1,s_2,s_3}\,\xi_{s_1-1}^{(1)}\varphi_{s_2}^\prime\varphi_{s_3}^{\prime(1)}\\\nonumber
& \hspace*{2cm}-\frac14\,(d-2+2Y_1\pl_{Y_1})\pl_{Y_3}^2\pl_{Y_1}f^{TT}_{s_1,s_2,s_3}\,\xi^{(1)}_{s_1-1}\,\Box_{2}\varphi_{s_2}\varphi_{s_3}^\prime
\end{align}
This requires: 
\begin{align}
    {\cal V}^{(3)}_{s_1,s_2,s_3}&=\frac18\,(d+2{\cal Y}_1\pl_{{\cal Y}_1})(d+2{\cal Y}_2\pl_{{\cal Y}_2})(d+2{\cal Y}_3\pl_{{\cal Y}_3})\pl_{{\cal Y}_1}^2\pl_{{\cal Y}_2}^2\pl_{{\cal Y}_3}^2f\,\varphi_{s_1}^{\prime(1)}\varphi_{s_2}^{\prime(1)}\varphi_{s_3}^{\prime(1)}\\\nonumber
    &+\frac18(d+2{\cal Y}_3\pl_{{\cal Y}_3})(d-2+2{\cal Y}_1\pl_{{\cal Y}_1})\pl_{{\cal Y}_1}\pl_{{\cal Y}_2}^2\pl_{{\cal Y}_3}^2\,U_1\cdot X_1\,\varphi_{s_1}^{\prime(2)}\varphi_{s_2}^\prime\varphi_{s_3}^{\prime(1)}+\text{cyclic}\\\nonumber
    &-\frac12\,\pl_{{\cal Y}_1}\pl_{{\cal Y}_2}\pl_{{\cal Y}_3}f\,U_1\cdot X_1\,\varphi_{s_1}^{\prime(1)}\,U_2\cdot X_2\,\varphi_{s_2}^{\prime(1)}\,U_3\cdot X_3\,\varphi_{s_3}^{\prime(1)},
\end{align}
which gives the complete cubic coupling \eqref{offshell0} in de Donder gauge. 

The variation is given by: 
\begin{equation}\label{d0v123}
\delta^{(0)}_{\xi_{s_1-1}}{\cal V}_{s_1,s_2,s_3}= {\hat T}_{13}{\hat { \cal F}}_{s_2}\varphi_{s_2}+{\hat T}_{12}{\hat { \cal F}}_{s_3}\varphi_{s_3},
\end{equation}
which is proportional to the equation of motion \eqref{dedondeomamb} for Fronsdal fields in de Donder gauge and thus vanishes on-shell, as required. The operators ${\hat T}_{12}$ and ${\hat T}_{13}$ finally read:
\begin{align}\label{F2}
{\hat T}_{13}\left(\xi_1,\varphi_{s_3}\right)=&-\frac12\pl_{{\cal Y}_1}f^{TT}_{s_1,s_2,s_3}\,\xi_1\varphi_{s_3}-\frac14(d-2+2{\cal Y}_1\pl_{{\cal Y}_1})\pl_{{\cal Y}_1}\pl_{{\cal Y}_3}^2f^{TT}_{s_1,s_2,s_3} \xi_1^{(1)}\varphi_{s_3}^{\prime}\\ \nonumber & -\frac14(d-2+2{\cal Y}_3\pl_{{\cal Y}_3})\pl_{{\cal Y}_1}\pl_{{\cal Y}_2}^2\pl_{{\cal Y}_3}f^{TT}_{s_1,s_2,s_3}\partial_{U_2}\cdot \partial_{U_2}\xi_1\,U_3\cdot X_3\,\varphi_{s_3}^{\prime(2)}\\  \nonumber &-\frac18(d+2{\cal Y}_3\pl_{{\cal Y}_3})(d-2+2{\cal Y}_1\pl_{{\cal Y}_1})\pl_{{\cal Y}_1}\pl_{{\cal Y}_2}^2\pl_{{\cal Y}_3}^2f^{TT}_{s_1,s_2,s_3}\,\partial_{U_2}\cdot \partial_{U_2} \xi_1^{(1)}\varphi_{s_3}^{\prime(1)},
\end{align}
and
\begin{align}\label{F3}
{\hat T}_{12}\left(\xi_1,\varphi_{s_2}\right)=&+\frac12\pl_{{\cal Y}_1}f^{TT}_{s_1,s_2,s_3}\,\xi_1\varphi_{s_2} +\frac14(d-2+2{\cal Y}_1\pl_{{\cal Y}_1})\pl_{{\cal Y}_1}\pl_{{\cal Y}_3}^2f^{TT}_{s_1,s_2,s_3}\partial_{U_3}\cdot \partial_{U_3} \xi_1^{(1)}\varphi_{s_2} \\ \nonumber
& -\frac14(d-2+2{\cal Y}_1\pl_{{\cal Y}_1})\pl_{{\cal Y}_1}\pl_{{\cal Y}_2}\pl_{{\cal Y}_3}^2f^{TT}_{s_1,s_2,s_3}\partial_{U_3}\cdot \partial_{U_3}\,\xi_1^{(1)} \,U_2\cdot X_2\,\varphi_{s_2}^{\prime(1)}\\ \nonumber
&-\frac18(d+2{\cal Y}_3\pl_{{\cal Y}_3})(d-2+2{\cal Y}_1\pl_{{\cal Y}_1})\pl_{{\cal Y}_1}\pl_{{\cal Y}_2}^2\pl_{{\cal Y}_3}^2f^{TT}_{s_1,s_2,s_3}\partial_{U_3}\cdot \partial_{U_3}\,\xi_1^{(1)} \varphi_{s_2}^{\prime(1)}\, \\ \nonumber
&-\frac12\pl_{{\cal Y}_1}\pl_{{\cal Y}_2}f^{TT}_{s_1,s_2,s_3}\,\xi_1\,U_2\cdot X_2\,\varphi_{s_2}^{\prime(1)},
\end{align}
Establishing the results of this section made extensive use of the following identity that holds for generic functions $f\left({\cal Y}_i\right)$ of ${\cal Y}_i$:\footnote{Where we introduced the operator $Q=-\tfrac{1}{2}\sum\limits^3_{i=1}\left(X_i \cdot \partial_{X_i}+U_i \cdot \partial_{U_i}\right)$.}
\begin{multline}
f({\cal Y}_i)\,\left(U_1\cdot X_1\right)=\left[(X_2 \cdot \partial_{X_2}+ U_2 \cdot \partial_{U_2}+Q-1-\tfrac{d}{2})\pl_{{\cal Y}_1}\right.\\\left.-\partial_{U_3} \cdot \partial_{X_3} \pl_{{\cal Y}_3}\pl_{{\cal Y}_1}-\partial_{U_3} \cdot \partial_{X_3}(Q+1-\tfrac{d}{2}+{\cal Y}_i\pl_{Y_i})\pl_{{\cal Y}_1}\pl_{{\cal Y}_3}^2\right]f({\cal Y}_i)\\ +(Q-\tfrac{d}{2}+{\cal Y}_i\pl_{{\cal Y}_i})\pl_{{\cal Y}_1}f\,X_1^2 \,,
\end{multline}
and the identities \eqref{idmanipn}.

\subsection{Ghost cubic couplings}
\label{subsec::gcc}

Using the standard Faddeev-Popov procedure, ghost fields are introduced upon exponentiating the determinant in the gauge fixed path integral:
\begin{equation}
\int[d\varphi]\,\Det\left(\frac{\delta}{\delta\xi}{\hat {\cal D}}(\delta_\xi\varphi)\right)\,e^{-S[\varphi]}=\int[d\varphi][d{\bar c}][d c]\,e^{-S[\varphi]-S_{\text{ghost}}\left[\varphi,{\bar c},c\right]}
\end{equation}
where $S[\varphi]$ is the action for the type A theory with ghost action 
\begin{equation}\label{ghostaction}
S_{\text{ghost}}\left[\varphi,{\bar c},c\right]=\sum_s(s-1)!\int_{\text{AdS}_{d+1}}\,{\bar c}_{s-1}\left(x,\partial_u\right)\frac{\delta}{\delta\xi_{s-1}}{\hat {\cal D}}(\delta_\xi\varphi)\,c_{s-1}\left(x,u\right).
\end{equation}
In the above we used $\varphi$ to collectively denote the Fronsdal fields in the spectrum of the type A higher-spin gauge theory, subject to the de Donder condition \eqref{dedonder}. The ${\bar c}$, $c$ collectively denote the corresponding ghost fields, which are traceless owing to the tracelessness of Fronsdal gauge parameters \eqref{gaugetransf}.

The ghost action \eqref{ghostaction} is entirely specified by the non-linear gauge transformations $\delta_{\xi}$ of the Fronsdal fields. In particular, expanding \eqref{ghostaction} up to cubic order
\begin{subequations}
{\small \begin{align}
 S_{\text{ghost}}\left[\varphi,{\bar c},c\right]&=S^{(2)}_{\text{ghost}}\left[\varphi,{\bar c},c\right]+S^{(3)}_{\text{ghost}}\left[\varphi,{\bar c},c\right]+...\,\\ \label{ghostfree}
S^{(2)}_{\text{ghost}}\left[\varphi,{\bar c},c\right]&=\sum^\infty_{s=2}(s-1)!\int_{\text{AdS}_{d+1}}\,
\bar{c}_{s-1}(x,\pl_{u})\,[\Box+\Lambda\,u\cdot\pl_u(u\cdot\pl_u+d-1)]\,c_{s-1}(x,u)\Big|_{u=0}\\
S^{(3)}_{\text{ghost}}\left[\varphi,{\bar c},c\right]&=\sum^\infty_{s=2}(s-1)!\int_{\text{AdS}_{d+1}}\, \bar{c}_{s-1}(x,\pl_u)\frac{\delta}{\delta \xi_{s-1}}{\hat {\cal D}}(\delta^{(1)}_{\xi_{s-1}}\varphi)\,c_{s-1}\left(x,u\right)\Big|_{u=0}\,,\label{ghostcubic}
\end{align}}
\end{subequations}
we see that the cubic vertices of the ghost action are determined by the first order deformation $\delta^{(1)}_{\xi}$ of the gauge transformations of the Fronsdal fields in de Donder gauge. Given the cubic action \eqref{osaction} for the type A theory and its off-shell variation \eqref{d0v123}, the the first order deformation $\delta^{(1)}_{\xi}$ is fixed by the cubic consistency condition \eqref{cubcinoe}:

 From the transformation of the free type A Fronsdal action \eqref{typeafree} under $\delta^{(1)}_{\xi_{s_1-1}}$ 
\begin{equation}
\delta^{(1)}_{\xi_{s_1-1}}S^{(2)}\left[\varphi\right]=\sum^\infty_{s=0}s!\int_{\text{AdS}_{d+1}} \delta^{(1)}_{\xi_{s_1-1}}\varphi_{s}\left(x,u\right)\left(1-\frac{1}{4}u^2 \partial_u \cdot \partial_u\right){\hat {\cal F}}_{s}\, \varphi_{s}\left(x,u\right).
\end{equation}
Combined with \eqref{d0v123}, the cubic consistency condition \eqref{cubcinoe} requires: 
\begin{subequations}\label{1odef}
\begin{align}
\delta^{(1)}_{\xi_{s_1-1}}\varphi_{s_2}&=-\frac{1}{s_2!}\left(1-\frac{1}{4}u^2_2\partial_{u_2} \cdot \partial_{u_2}\right)^{-1}{\hat T}_{13}\left(\xi_1,\varphi_{s_3}\right), \\
\delta^{(1)}_{\xi_{s_1-1}}\varphi_{s_3}&=-\frac{1}{s_3!}\left(1-\frac{1}{4}u^2_3\partial_{u_3} \cdot \partial_{u_3}\right)^{-1}{\hat T}_{12}\left(\xi_1,\varphi_{s_2}\right),
\end{align}
\end{subequations}
where ${\hat T}_{13}$ and ${\hat T}_{12}$ were defined in the linearised variation \eqref{d0v123} of $S^{(3)}$. 

For the ghost cubic action \eqref{ghostcubic}, we thus have
\begin{align}
S^{(3)}_{\text{ghost}}\left[\varphi,{\bar c},c\right]&=-\sum^\infty_{s_2=2}\frac{1}{s_2}\int_{\text{AdS}_{d+1}}\, \bar{c}_{s_2-1}(x,\pl_{u_2}) \left(\partial_{u_2}\cdot \nabla_2\right){\hat T}_{13}\left(c_{s_1-1},\varphi_{s_3}\right)\Big|_{u=0}\\&-\sum^\infty_{s_3=2}\frac{1}{s_3}\int_{\text{AdS}_{d+1}}\, \bar{c}_{s_3-1}(x,\pl_{u_3}) \left(\partial_{u_3}\cdot \nabla_3\right){\hat T}_{12}\left(c_{s_1-1},\varphi_{s_2}\right)\Big|_{u=0},\nonumber
\end{align}
where we used the identity:
\begin{equation}\label{deDonId}
\mathcal{D}\left(1-\frac14u^2\pl_u^2\right)^{-1}=\pl_u\cdot\nabla+{\cal O}(u^2)\,.
\end{equation}
The ${\cal O}(u^2)$ terms do not contribute owing to the tracelessness of the ghost fields. 

\section{Propagators}

\subsection{Bulk-to-boundary propagators}

\subsubsection{Review: Fronsdal bulk-to-boundary propagators}

The solution of the source-free Fronsdal equation \eqref{eomfronsdal} in de Donder gauge 
\begin{equation}\label{sfrededoneq}
\left(\Box- m^2_s-u^2\partial_u\cdot \partial_u\right)\varphi_s\left(x,u\right) = 0,
\end{equation}
subject to the standard AdS/CFT boundary condition for spin-$s$ gauge fields\footnote{For concreteness, here we used co-ordinates $x^\mu=\left(y^i,r\right)$ for AdS$_{d+1}$:
\begin{align}
   ds^2 = e^{2r}dy_idy^i+R^2dr^2,
\end{align}
with $R$ the AdS radius. The boundary of AdS is located at $r=\infty$, with boundary directions $y^i$, $i = 1, ..., d$.}
\begin{equation}
    \lim_{r \rightarrow \infty} \varphi_{\mu_1...\mu_s}\left(y,r\right)e^{2\left(1-s\right)r} \: = \: {\bar \varphi}_{i_1...i_s}\left(y\right) \label{bcbubo},
\end{equation}
can be constructed from the bulk-to-boundary propagator:
\begin{equation}
\varphi_s\left(x,u\right)=\int_{\partial\text{AdS}_{d+1}}d^dy^\prime\, K_{s}\left(x,u;y^\prime,{\hat \partial}_z\right) {\bar \varphi}_s\left(y^\prime,z\right),
\end{equation}
where:
\begin{subequations}\label{bubos}
\begin{align}\label{bcbubos}
\left(\Box- m^2_s-u^2\partial_u\cdot \partial_u\right) K_{s}\left(x,u;y^\prime,z\right)&=0,\\ \label{bcbuboslim}
\lim_{r \rightarrow \infty}\left(e^{2\left(1-s\right)r}K_{\mu_1 ... \mu_s}{}^{i_1 ... i_s}\left(y,r;y^\prime\right)\right) & = \frac{\delta{}^{i_1 \, \ldots}_{\left\{\right.\mu_1 \, \ldots}\delta{}^{i_s }_{\mu_s\left.\right\}}}{2s+d-2}\delta^{d}\left(y-y^\prime\right).
\end{align}
\end{subequations}
Since the equation \eqref{sfrededoneq} for the bulk-to-boundary propagator is source-free, we can go on-shell and choose the traceless and tranverse gauge \eqref{ttintconds} using the on-shell gauge parameters \eqref{dedondergaugetran}. The equation of motion for the propagator becomes:
\begin{subequations}
\begin{align}
\left(\Box- m^2_s\right) K_{s}\left(x,u;y^\prime,z\right)&=0,\\
\left(\partial_u \cdot \nabla\right)K_{s}\left(x,u;y^\prime,z\right)&=0,\\
\left(\partial_u \cdot \partial_u\right)K_{s}\left(x,u;y^\prime,z\right)&=0,
\end{align}
\end{subequations}
with boundary condition \eqref{bcbuboslim} unchanged. It is most straightforward to solve for the bulk-to-boundary propagator in ambient space, in which the ambient representative of the bulk-to-boundary propagator satisfies
\begin{subequations}\label{ambeom}
\begin{align}
\partial^2_X K_s\left(X,U;P,Z\right)&=0, \\
\left(\partial_U \cdot \partial_X \right)K_s\left(X,U;P,Z\right)&=0, \\
\left(\partial_U \cdot \partial_U \right)K_s\left(X,U;P,Z\right)&=0.
\end{align}
\end{subequations}
Together with the tangentiality and homogeneity conditions:
\begin{subequations}
\begin{align}
\left(X \cdot \partial_U \right)K_s\left(X,U;P,Z\right) = \left(X \cdot \partial_X+s+d-2\right)K_s\left(X,U;P,Z\right)&=0,\\
\left(Z \cdot \partial_P\right)K_s\left(X,U;P,Z\right)=\left(P \cdot \partial_P+s+d-2\right)K_s\left(X,U;P,Z\right)&=0,
\end{align}
\end{subequations}
the solution is fixed uniquely up to an overall coefficient \cite{Mikhailov:2002bp}:
\begin{align}\label{bubospins}
    K_{s}\left(X, U; P, Z \right) & = C_{s+d-2,s}\frac{\left[\left(U \cdot Z\right)\left(-2P \cdot X\right)+2\left(U \cdot P\right)\left(Z \cdot X\right)\right]^s}{\left(-2 X \cdot P\right)^{2s+d-2}}. 
\end{align}
The coefficient is fixed by equation \eqref{bcbuboslim} to be:
\begin{align} \label{bubonorm}
    C_{s+d-2,s}& = \frac{\left(2s+d-3\right)\Gamma\left(s+d-2\right)}{2 \pi^{d/2} \left(s+d-3\right) \Gamma\left(s-1+\tfrac{d}{2}\right)}.
\end{align}

In ambient space it is straightforward to extend the above result to the bulk-to-boundary propagator $K_{\Delta,s}$ of a totally symmetric spin-$s$ field of generic mass $m^2R^2=\Delta\left(\Delta-d\right)-s$, which has the same ambient equation of motion \eqref{ambeom} but with the homogeneity degree:
\begin{equation}
\left(X \cdot \partial_X+\Delta\right)K_{\Delta,s}\left(X,U;P,Z\right)=0.
\end{equation}
The result is simply \cite{Mikhailov:2002bp,Costa:2014kfa}: 
\begin{subequations}\label{bubospinsgencoef}
\begin{align}\label{bubospinsgen}
    K_{\Delta,s}\left(X, U; P, Z \right) & = C_{\Delta,s}\frac{\left[\left(U \cdot Z\right)\left(-2P \cdot X\right)+2\left(U \cdot P\right)\left(Z \cdot X\right)\right]^s}{\left(-2 X \cdot P\right)^{\Delta+s}},\\ \label{bubonormgen}
    C_{\Delta,s}& = \frac{\left(s+\Delta-1\right)\Gamma\left(\Delta\right)}{2 \pi^{d/2} \left(\Delta-1\right) \Gamma\left(\Delta+1-\tfrac{d}{2}\right)},
\end{align}
\end{subequations}
which coincides with \eqref{bubospins} for $\Delta=s+d-2$, as required.

\subsubsection{Ghost boundary-to-bulk propagators}
\label{subsubsec:ghbobu}

The bulk-to-boundary propagator for the ghost associated to a spin-$s$ gauge field is the solution to the Fierz system:
\begin{subequations}
\begin{align}
\left(-\Box+m^2_\xi \right)K^{\text{gh.}}_{s-1}\left(x,u;y,z\right)&=0, \\
\left(\partial_u \cdot \partial_u\right)K^{\text{gh.}}_{s-1}\left(x,u;y,z\right)&=0,\\
\left(\partial_u \cdot \nabla \right)K^{\text{gh.}}_{s-1}\left(x,u;y,z\right)&=0,
\end{align}
\end{subequations}
which is represented in ambient space as 
\begin{subequations}\label{amghheom}
\begin{align}\label{eomghbubo}
    \left[\partial^2_X+\frac{2}{X^2}(d-2+2U\cdot\pl_U)\right]K^{\text{gh.}}_{s-1}(X,U;P,Z)&=0\,,\\
    \pl_{U}\cdot\pl_{X}\,K^{\text{gh.}}_{s-1}(X,U;P,Z)&=0\,,\\
    \pl_{U}^2\,K^{\text{gh.}}_{s-1}(X,U;P,Z)&=0\,,
\end{align}
\end{subequations}
subject to the following homogeneity and tangentiality conditions
\begin{subequations}\label{homobubogh}
\begin{align}
     \left(X\cdot\pl_U\right)\,K^{\text{gh.}}_{s-1}(X,U;P,Z)&=(X\cdot\pl_{X}+s+d-3)K^{\text{gh.}}_{s-1}(X,U;P,Z)=0,\\
     \left(P\cdot\pl_Z\right)\,K^{\text{gh.}}_{s-1}(X,U;P,Z)&=(P\cdot\pl_{P}+s+d-3)K^{\text{gh.}}_{s-1}(X,U;P,Z)=0.
\end{align}
\end{subequations}
By virtue of the commutator \eqref{app::amsscomm}, the solution for the propagator can be obtained by dressing with $X^2$ the solution to the \emph{massless} ambient space Fierz system \eqref{ambeom} of the previous section, to accommodate for the non-zero mass term in the ambient equation of motion \eqref{eomghbubo}. To wit, we make an ansatz of the form
\begin{equation}
K^{\text{gh.}}_{s-1}(X,U;P,Z)=\frac{1}{\left(X^2\right)^\alpha}K_{d+s-3-2\alpha,s-1}(X,U;P,Z),
\end{equation}
where $K_{d+s-3-2\alpha,s-1}$ is given by \eqref{bubospinsgencoef} with $\Delta=d+s-3-2\alpha$. The $-2\alpha$ is to preserve the homogeneity degree \eqref{homobubogh}. Plugging the ansatz into the equation of motion \eqref{amghheom}, one finds that $\alpha=-1$. 

To summarise, the bulk-to-boundary propagator for a ghost field associated to a spin $s$ gauge field is given in ambient space by
\begin{equation}
K^{\text{gh.}}_{s-1}(X,U;P,Z)=C_{d+s-1,s-1}X^2\frac{\left[\left(U \cdot Z\right)\left(-2P \cdot X\right)+2\left(U \cdot P\right)\left(Z \cdot X\right)\right]^{s-1}}{\left(-2 X \cdot P\right)^{d+2s-2}},
\end{equation}
where $C_{d+s-1,s-1}$ is defined in \eqref{bubonormgen}.

\subsection{Bulk-to-bulk propagators}
\label{subsec::bubuprop}

\subsubsection{Bulk-to-bulk propagators of Fronsdal fields in de-Donder gauge}

The solution to the Euler-Lagrange equations for the Fronsdal action \eqref{Fronsdalaction} with some source $J_s$,\footnote{The double-traceless condition on the source $J_s$ arises from the double-tracelessness of Fronsdal fields.}
\begin{equation}
\left(1-\frac{1}{4}u^2_1 \partial_{u_1}\cdot \partial_{u_1}\right){\hat {\cal F}}_s \varphi_s\left(x,u\right) = -J_s\left(x,u\right), \qquad \left(\partial_u \cdot \partial_u\right)^2J_{s}\left(x,u\right)=0,
\end{equation}
can be expressed in terms of the bulk-to-bulk propagator $\Pi_s$, 
\begin{equation}
\varphi_s\left(x,u\right)=s! \int_{\text{AdS}_{d+1}} \Pi_s\left(x_1,u_1;x_2,\partial_{u_2}\right)J_s\left(x_2,u_2\right),
\end{equation}
which satisfies the simpler equation 
\begin{multline}\label{fronbbpropeq}
\left(1-\frac{1}{4}u^2_1 \partial_{u_1}\cdot \partial_{u_1}\right)
{\hat {\cal F}}_{s}\left(x_1,u_1\right)
 \Pi_s\left(x_1,u_1;x_2,u_2\right) \\ = -\left\{\left\{\left(u_1 \cdot u_2\right)^s \delta^{d+1}\left(x_1,x_2\right)\right\}\right\}+\left(u_2 \cdot \nabla_2\right)\Lambda_{s,s-1}\left(x_1,u_1;x_2,u_2\right),
\end{multline}
where $\Lambda_{s,s-1}$ is a pure gauge term, subject to the constraints
\begin{equation}
\left(\partial_{u_1}\cdot \partial_{u_1} \right)^2\Lambda_{s,s-1}\left(x_1,u_1;x_2,u_2\right)=\left(\partial_{u_2}\cdot \partial_{u_2} \right)\Lambda_{s,s-1}\left(x_1,u_1;x_2,u_2\right)=0.
\end{equation}
The pure gauge term $\Lambda_{s,s-1}$ is is often disregarded  when the source $J_s$ is conserved.\footnote{To be more precise, in the case of Fronsdal fields, $\Lambda_{s,s-1}$ drops out if the source is conserved up to traces: $\left(\partial_u \cdot \nabla\right) J_s\left(x,u\right)={\cal O}\left(u^2\right)$. This is owing to the tracelessness of the Fronsdal gauge parameters.} In the context of Witten diagrams, this is the case when the constituent fields of $J_s$ are external and thus on-shell. See \cite{Francia:2007qt,Francia:2008hd,Manvelyan:2008ks,Bekaert:2014cea,Bekaert:2015tva,Sleight:2016hyl,Sleight:2017fpc} in the context of the four-point tree-level exchange of a spin-$s$ gauge field. However when the source is off-shell, such as in loop diagrams, this no longer holds and the explicit form of $\Lambda_{s,s-1}$ is required.

In the following we complete the results of \cite{Francia:2007qt,Francia:2008hd,Manvelyan:2008ks,Bekaert:2014cea} and determine the full spin-$s$ bulk-to-bulk propagator \eqref{fronbbpropeq} in de Donder gauge, including the form of the pure gauge terms $\Lambda_{s,s-1}$. To this end, it is useful to express the double-traceless Fronsdal field in terms of its irreducible components \eqref{irredecomp}, for which it is useful to employ the notation
\begin{equation}
    \varphi_s(x,u)=\begin{pmatrix}
    {\tilde \varphi}_{s}(x,w)\\
    \varphi^\prime_s (x,w)
    \end{pmatrix},
\end{equation}
where we have introduced a traceless auxiliary variable $w$ with the property $w^2=0$, which ensures the tracelessness of each component. See \S \tcb{\ref{sec::notconvamb}}. In this notation, the Fronsdal bulk-to-bulk propagator is thus a $2 \times 2$ matrix
\begin{equation}\label{2x2matrix}
    \Pi_s(x_1,u_1;x_2,u_2)=\begin{pmatrix}
    \pi_{{\tilde \varphi}_1{\tilde \varphi}_2}&\pi_{{\tilde \varphi}_1\varphi_2^\prime} \\
    \pi_{\varphi_1^\prime{\tilde \varphi}_2}&\pi_{\varphi_1^\prime\varphi_2^\prime}
    \end{pmatrix}\,,
\end{equation}
which involves off-diagonal mixing terms between the  two off-shell irreducible components of a doubly traceless Fronsdal field.

In the de Donder gauge, the gauge fixed equation of motion for the bulk-to-bulk propagator is\footnote{Where for convenience we re-defined $\Lambda_{s,s-1} \rightarrow \left(1-\frac{1}{4}u^2_1 \partial_{u_1} \cdot \partial_{u_1}\right)\Lambda_{s,s-1}$.}
\begin{multline}
\label{FronsdaltensorDD}
(1-\frac{1}{4} \,u^2_1\, \partial_{u_1} \cdot \partial_{u_1})\left[(\Box_1- m^2_s)-u^2_1(\partial_{u_1}\cdot \partial_{u_1})\right]\Pi_{s}(x_1,u_1;x_2,u_2)\\
= -\left\{\left\{\left(u_1\cdot u_2\right)^s\delta^{d+1}\left(x_1,x_2\right)\right\}\right\}+\left(u_2\cdot\nabla_2\right)(1-\frac{1}{4} \,u^2_1\, \partial_{u_1} \cdot \partial_{u_1})\Lambda_{s,s-1}\left(x_1,u_1;x_2,u_2\right),
\end{multline}
and propagator must also satisfy the de Donder condition:
\begin{equation}\label{dedoncondprop}
{\hat {\cal D}}_1\Pi_{s}(x_1,u_1;x_2,u_2)=\left(\nabla_1\cdot{\pl}_{u_1}-\tfrac12\,u_1\cdot\nabla\pl_{u_1}^2\right)\Pi_{s}(x_1,u_1;x_2,u_2)=0\,.
\end{equation}
To solve for the propagator it is most straightforward to use ambient space with constrained auxiliary variables \eqref{trauxamb}. In terms of the latter, and furthermore focusing on the irreducible components \eqref{2x2matrix}, the de Donder condition \eqref{dedoncondprop} reads: 

{\small \begin{align}\label{matrdedoncond}
\begin{pmatrix}
\nabla_1\cdot\hat{D}_{W_1}\pi_{{\tilde \varphi}_1{\tilde \varphi}_2}+\left(\tfrac1{2s+d-3}-\tfrac12\right)W_1\cdot\nabla_1\,\pi_{\varphi_1^\prime{\tilde \varphi}_2}& \nabla_1\cdot\hat{D}_{W_1}\pi_{{\tilde \varphi}_1\varphi_2^\prime}+\left(\tfrac1{2s+d-3}-\tfrac12\right)W_1\cdot\nabla_1\,\pi_{\varphi_1^\prime\varphi_2^\prime} \\
0&0
\end{pmatrix}=0_{2 \times 2}\,,
\end{align}}

\noindent
where the zeros on the second line are owing to the tracelessness of the de-Donder operator. To solve for the propagator, we can decompose the four trace-less components in the basis \S \tcb{\ref{app::adsharm}} of bi-tensorial harmonic functions $\Omega$ on AdS$_{d+1}$:

{\small \begin{subequations}
\begin{align}\label{eq:ManifestTrace}
\pi_{{\tilde \varphi}_1{\tilde \varphi}_2}(X_1,W_1;X_2,W_2) &= \sum^{s}_{l=0} \int^{\infty}_{-\infty}  d\nu \; g^{{\tilde \varphi}_1{\tilde \varphi}_2}_{s,s-l}\left(\nu\right) \left(W_1\cdot\nabla\right)^{l} \left(W_2\cdot\nabla\right)^{l} \Omega_{\nu,s-l}(X_1,W_1;X_2,W_2),\\
\pi_{{\tilde \varphi}_1\varphi^\prime_2}(X_1,W_1;X_2,W_2) &= \sum^{s-2}_{l=0} \int^{\infty}_{-\infty}  d\nu \; g^{{\tilde \varphi}_1\varphi^\prime_2}_{s,s-l-2}\left(\nu\right) \left(W_1\cdot\nabla\right)^{l+2} \left(W_2\cdot\nabla\right)^{l} \Omega_{\nu,s-2-l}(X_1,W_1;X_2,W_2),\\
\pi_{\varphi^\prime_1{\tilde \varphi}_2}(X_1,W_1;X_2,W_2) &= \sum^{s-2}_{l=0} \int^{\infty}_{-\infty}  d\nu \; g^{\varphi^\prime_1{\tilde \varphi}_2}_{s,s-l-2}\left(\nu\right) \left(W_1\cdot\nabla\right)^{l} \left(W_2\cdot\nabla\right)^{l+2} \Omega_{\nu,s-2-l}(X_1,W_1;X_2,W_2),\\
\pi_{\varphi^\prime_1\varphi^\prime_2}(X_1,W_1;X_2,W_2) &= \sum^{s-2}_{l=0} \int^{\infty}_{-\infty}  d\nu \; g^{\varphi^\prime_1\varphi^\prime_2}_{s,s-l-2}\left(\nu\right) \left(W_1\cdot\nabla\right)^{l} \left(W_2\cdot\nabla\right)^{l} \Omega_{\nu,s-2-l}(X_1,W_1;X_2,W_2),
\end{align}
\end{subequations}}

\noindent
where the functions $g^{{\tilde \varphi}_1{\tilde \varphi}_2}_{s,s-l}\left(\nu\right)$, $g^{{\tilde \varphi}_1\varphi^\prime_2}_{s,s-l-2}\left(\nu\right)$, $g^{\varphi^\prime_1{\tilde \varphi}_2}_{s,s-l-2}\left(\nu\right)$ and $g^{\varphi^\prime_1\varphi^\prime_2}_{s,s-l-2}\left(\nu\right)$ are to be determined. 

The symmetry under $\left(x_1,u_1\right) \leftrightarrow \left(x_2,u_2\right)$ requires $g^{{\tilde \varphi}_1\varphi^\prime_2}_{s,s-l-2}\left(\nu\right)=g^{\varphi^\prime_1{\tilde \varphi}_2}_{s,s-l-2}\left(\nu\right)$, while the de Donder condition \eqref{matrdedoncond} demands that: 
\begin{subequations}\label{dedonderrela}
\begin{align}
g^{{\tilde \varphi}_1\varphi^\prime_2}_{s,s-l-2}\left(\nu\right) &= \frac{(l+2) (-d+l-2 s+4) \left((d+2 s-2)^2+4 \nu ^2\right)}{2 (d+2 s-5)}g^{{\tilde \varphi}_1{\tilde \varphi}_2}_{s,s-l-2}\left(\nu\right)\\
g^{\varphi^\prime_1\varphi^\prime_2}_{s,s-l-2}\left(\nu\right)&= \frac{(l+2) (-d+l-2 s+4) \left((d+2 s-2)^2+4 \nu ^2\right)}{2 (d+2 s-5)}g^{{\tilde \varphi}_1\varphi^\prime_2}_{s,s-l-2}\left(\nu\right).
\end{align}
\end{subequations}
The above relations can be straightforwardly derived using \eqref{divgradmarm}.

With the de Donder condition fulfilled, the propagator and the gauge term $\Lambda_{s,s-1}$ are fixed uniquely by the equation of motion \eqref{FronsdaltensorDD}. The latter can be likewise expanded in the basis of harmonic functions, as:
\begin{equation}
 \Lambda_{s,s-1} = \left(
\begin{array}{cc}
\Lambda_{\tilde \varphi_1}  & 0 \\
\Lambda_{\varphi^\prime_1} & 0 \\
\end{array}
\right)
\end{equation}
where 
\begin{subequations}\label{gtansaetze}
\begin{align}
\Lambda_{\tilde \varphi_1}(X_1,W_1;X_2,W_2)&=\sum^{s}_{l=0} \int^{\infty}_{-\infty}  d\nu \; \lambda^{{\tilde \varphi}_1}_{s,s-l}\left(\nu\right) \left(W_1\cdot\nabla\right)^{l} \left(W_2\cdot\nabla\right)^{l-1} \Omega_{\nu,s-l}(X_1,W_1;X_2,W_2) \\
\Lambda_{\varphi^\prime_1}(X_1,W_1;X_2,W_2)&=\sum^{s-2}_{l=0} \int^{\infty}_{-\infty}  d\nu \; \lambda^{\varphi^\prime_1}_{s,s-l-2}\left(\nu\right) \left(W_1\cdot\nabla\right)^{l} \left(W_2\cdot\nabla\right)^{l+1} \Omega_{\nu,s-l-2}(X_1,W_1;X_2,W_2).
\end{align}
\end{subequations}
Plugging in the propagator ansatz \eqref{eq:ManifestTrace2} with the de Donder constraints \eqref{dedonderrela} and the above ans\"atze for the gauge terms \eqref{gtansaetze} into the equation of motion \eqref{FronsdaltensorDD}, we find:
\begin{subequations}
 \begin{align}
g^{{\tilde \varphi}_1{\tilde \varphi}_2}_{s,s-l}\left(\nu\right)&=
\frac{64s(s-1)(d+2 s-5)}{l (-d+l-2 s+2)\left((d+2 s-2)^2+4 \nu ^2\right)} \\ \nonumber &  \hspace*{1.5cm} \times
\frac{c^{(s-2)}_{l-2}\left(\nu\right)}{ \left(d^2+4 d (l+s-2)+4 \left(-l^2+2 l (s-1)+\nu ^2+(s-2)^2\right)\right)^2}\\
g^{{\tilde \varphi}_1\varphi^\prime_2}_{s,s-2-l}\left(\nu\right)&=\frac{(l+2) (-d+l-2 s+4) \left((d+2 s-2)^2+4 \nu ^2\right) }{2 (d+2 s-5)}\,g^{{\tilde \varphi}_1{\tilde \varphi}_2}_{s,s-2-l}\left(\nu\right)\,,\\
g^{\varphi^\prime_1{\tilde \varphi}_2}_{s,s-2-l}\left(\nu\right)&=\frac{(l+2) (-d+l-2 s+4) \left((d+2 s-2)^2+4 \nu ^2\right) }{2 (d+2 s-5)}\,g^{{\tilde \varphi}_1{\tilde \varphi}_2}_{s,s-2-l}\left(\nu\right)\,,\\
g^{\varphi^\prime_1\varphi^\prime_2}_{s,s-2-l}\left(\nu\right)&=\frac{(l+2)^2 (-d+l-2 s+4)^2 \left((d+2 s-2)^2+4 \nu ^2\right)^2}{4 (d+2 s-5)^2}g^{{\tilde \varphi}_1{\tilde \varphi}_2}_{s,s-2-l}\left(\nu\right)\,,
\end{align}
\end{subequations}
and, for the gauge term:
\begin{subequations}
\begin{align}
    \lambda^{{\tilde \varphi}_1}_{s,s-l}\left(\nu\right)&=(2s+5-d)k_{s,l}\left(\nu\right)\\  \lambda^{\varphi^\prime_1}_{s,s-l}\left(\nu\right)&=\frac{1}{2} (l-1) (-d+l-2 s+3) \left((d+2 s-4)^2+4 \nu ^2\right)k_{s,l}\left(\nu\right)\,,
\end{align}
\end{subequations}
where we have defined for convenience:
\begin{align}
k_{s,l}\left(\nu\right)&=-\frac{c^{(s)}_{l}\left(\nu\right)}{d+2 s-5}\\ \nonumber &-\frac{16 (s-1) s}{l (d-l+2 s-2) \left((d+2 s-2)^2+4 \nu ^2\right)}\\ \nonumber & \hspace*{3cm} \times
-\frac{c^{(s-2)}_{l-2}\left(\nu\right)}{\left(d^2+4 d (l+s-2)+4 \left(2 l s-l (l+2)+\nu ^2+s^2-4 s+4\right)\right)}
\end{align}
The above results are straightforward to obtain using \eqref{boxgradharm} and the completeness relation \eqref{eq:Completeness} where $c^{(s)}_{l}\left(\nu\right)$ is given explicitly by \eqref{csl}.

Note that in this section we determined the bulk-to-bulk propagator using the constrained ambient space auxiliary variables \eqref{trauxamb}. The unconstrained form can be obtained uniquely from the homogeneity and tangentiality conditions \eqref{homo} and \eqref{tangent}.

\subsubsection{Bulk-to-bulk propagators of ghost fields}

The equation for the bulk-to-bulk propagator of a ghost field $c_{s-1}$ associated to a spin-$s$ Fronsdal field in de Donder gauge can be derived from the Euler-Lagrangian equations of the free ghost action \eqref{ghostfree}. The resulting propagator equation is:
\begin{equation}
    (\Box-m^2_\xi)\Pi^{\text{gh.}}_{s-1}(x_1,w_1;x_2,w_2)=(w_1\cdot w_2)^{s-1}\delta^{d+1}(x_1,x_2)\,,
\end{equation}
where we recall that $m^2_{\xi} R^2=\left(s-1\right)\left(s+d-2\right)$. The ghost fields and thus their bulk-to-bulk propagators are traceless, so the harmonic function decomposition of the latter takes the form:
{\small 
\begin{equation}
\Pi^{\text{gh.}}_{s-1}(X_1,W_1;X_2,W_2) = \sum^{s-1}_{l=0} \int^\infty_{-\infty}d\nu\,h_{s-1,s-1-l}\left(\nu\right) \left(W_1\cdot\nabla\right)^{l} \left(W_2\cdot\nabla\right)^{l} \Omega_{\nu,s-1-l}(X_1,W_1;X_2,W_2). 
\end{equation}}
Using the identity \eqref{boxgradharm} and completeness relation \eqref{eq:Completeness}, the coefficients $h_{s-1,s-1-l}\left(\nu\right)$ can be straightforwardly determined:
\begin{equation}
h_{s-1,s-1-l}\left(\nu\right)=-c^{(s-1)}_{l}\left(\nu\right)\frac{d+2s-5}{\left(l-1\right)\left(2s+d-l-5\right)} \frac{1}{\nu^2+\left(s-3+\frac{d}{2}\right)^2}.
\end{equation}

\section{Beyond cubic order}

We conclude presenting technical trick which, given the gauge-fixed de Donder gauge vertices of Fronsdal fields, allows to obtain the corresponding ghost vertices directly in the ambient space formalism. This trick is shown to work at any order in the weak fields once the corresponding de-Donder completion of the vertices is obtained. 

The key observation is the identity:
\begin{equation}\label{deDonId2}
\mathcal{D}\left(1-\frac14u^2\pl_u^2\right)^{-1}=\pl_u\cdot\nabla+O(u^2)\,,
\end{equation}
from which, keeping track of normalisations, one obtains the following ghost vertex in terms of the variation of the physical vertex:
\begin{equation}
S_{\text{ghost}}^{(3)}=\frac1{s}\int_{AdS_{d+1}} e^{\pl_{u_1}\cdot\pl_{u_2}}\left[(u_1\cdot\nabla_1\,\bar{c}_{s-1})_{u_1}\, \left(\frac{\delta}{\delta\xi_{s-1}}T(\varphi,\xi)\,c_{s-1}\, \varphi\right)_{u_2}\right]\,.
\end{equation}
The observation is that the above ghost action is exactly proportional to the gauge variation of the de-Donder vertex up to a simple substitution, which can be identified as:
\begin{equation}
    f\tfrac1{X_3^2}\,c_1\,\varphi_2\,(X_3^2\mathcal{F}_3\varphi_3)\rightarrow \frac1{s_3}\,f\tfrac1{X_3^2}\,c_1\,\varphi_2\,(U_3\cdot\nabla_3 c_3)\,,
\end{equation}
where we have used that the degree of homogeneity of $U_3\cdot\nabla_3 c_3$ is the same as that of $\varphi$ together with the fact that both $U_3\cdot\nabla_3 c_3$ and $X_3^2\mathcal{F}_3\varphi_3$ are tangent tensors. In particular, we arrive to the following list of substitutions:
\begin{subequations}\label{ghostsub}
\begin{align}
(\mathcal{F}\varphi_s)^{(n)}&\rightarrow \frac1{s}\left(U\cdot \pl_{X}\,c^{(n+1)}+U\cdot X\,(d+2n+2U\cdot \pl_U)c^{(n+2)}\right)\,,\\
(\mathcal{F}\varphi_s)^{\prime(n)}&\rightarrow \frac{2}{s}\,\pl_{U}\cdot \pl_{X}\,c^{(n+1)}\,,
\end{align}
\end{subequations}
which trivially allow to recover the ghost vertex from \eqref{F2} and \eqref{F3}.

The latter substitutions generalise to all orders in the Noether procedure. Indeed the $n$-th order deformation of the gauge transformation has also the same general structure as above:
\begin{equation}
    \delta_{\xi}^{(n)}\varphi(u)=\frac{1}{s!}\left(1-\tfrac14 u^2\pl_{u}^2\right)^{-1}T^{(n)}(\xi,\varphi)\,.
\end{equation}
with the functional $T^{(n)}$ extracted from:
\begin{equation}
\delta_c^{(0)}S^{(n)}+\ldots+\delta_c^{(n-1)}S^{(3)}=T^{(n)}(c,\varphi)\mathcal{F}\varphi\,.
\end{equation}
One can then obtain the ghost vertex at order $n$ simply from the above right-hand side, with the substitutions \eqref{ghostsub}.

\subsection*{Acknowledgements}

C. S. and M. T. thank Simone Giombi for useful discussions and comments. The research of M. T. is partially supported by the Fund for Scientific Research-FNRS Belgium, grant FC 6369, the Russian Science Foundation grant 14-42-00047 in association with Lebedev Physical Institute and by the INFN within the program ``New Developments in AdS$_3$/CFT$_2$ Holography''. The research of C. S. was partially supported by the INFN and ACRI's (Associazione di Fondazioni e di Casse di Risparmio S.p.a.) Young Investigator Training Program, as part of the Galileo Galilei Institute for Theoretical Physics (GGI) workshop ``New Developments in AdS$_3$/CFT$_2$ Holography''.

\begin{appendix}
\section{Operator algebras}
\label{app::opalg}

\subsection{Intrinsic totally symmetric fields}
\label{app::opalg1}

\begin{subequations}
\begin{align}
[\nabla_\mu,\nabla_\nu]& =
		\Lambda(u_\mu\partial_{u_\nu}-u_\nu\partial_{u_\mu})\\
	[\Box,u\cdot\nabla]&=\Lambda \left[u\cdot\nabla(2u\cdot\partial_u+d-1)-2u^2\nabla\cdot\partial_u\right]\\
	[\nabla\cdot\partial_u,\Box]&=\Lambda\left[(2u\cdot\partial_u+d-1)\nabla\cdot\partial_u-2u\cdot \nabla \partial_u^2\right]\\
	[\nabla\cdot\partial_u,u\cdot\nabla]&=\Box+\Lambda\left[u\cdot\partial_u(u\cdot\partial_u+d-2)-u^2\partial_u^2\right]\\
	[\nabla\cdot\partial_u,u^2]&=2u\cdot\nabla,\\
	[\partial_u^2,u\cdot\nabla]&=2\nabla\cdot\partial_u,\\
	[\partial_u^2,u^2]&=2(d+2u\cdot\partial_u).
\end{align}
\end{subequations}

\subsection{Intrinsic traceless fields}
\label{app::opalg2}

\begin{subequations}\label{eq:operatoralgebra2}
\begin{align}
[\hat\partial_{w^\mu},w^\nu] & = 
		g_{\mu\nu}-\frac{2}{d-1+2 w\cdot \partial_w}w_\mu\hat\partial_{w^\mu},\\
[\nabla_\mu,\nabla_\nu]&=\Lambda(w_\mu\hat\partial_{w_\nu}-w_\nu\hat\partial_{w_\mu}),\\
[\Box,w\cdot\nabla]	&=\Lambda w\cdot\nabla(2w\cdot\hat\partial_u+d-1),\\
[\nabla\cdot\hat\partial_w,\Box]&=\Lambda(2w\cdot\hat\partial_w+d-1)\nabla\cdot\hat\partial_w\\
[\nabla\cdot\hat\partial_w,w\cdot\nabla]&=\Box-\frac{2}{d-1+2 w\cdot \partial_w}w\cdot\nabla\nabla\cdot\hat\partial_w+\Lambda w\cdot\hat\partial_w(w\cdot\hat\partial_w+d-2).
\end{align}
\end{subequations}

\subsection{Totally symmetric ambient fields}
\label{app::opalg3}

\begin{subequations}
\begin{align}
    [X\cdot\partial_U,\nabla_A]&=0\,,\\ [\partial_U\cdot\partial_U,\nabla_A]&=0\,,\\ [\nabla_A,X^2]&=0,\\ [D_{U}^A,\nabla^B]&=\tfrac{X^A}{X^2}\,D_U^B,
\end{align}
\end{subequations}
From which one can derive the useful identity:
\begin{equation}\label{app::amsscomm}
[\Box,(X^2)^{-n}]=-2n(2X\cdot\pl_X+d+2n+4)(X^2)^{-n-1}.
\end{equation}

It is also useful to note that:
\begin{subequations}\label{actcodamrep}
\begin{align}
\nabla\cdot\pl_U f\left(X,U\right)&=\pl_{U}\cdot\pl_X f\left(X,U\right)+\frac{U\cdot X}{X^2}\,\pl_{U}\cdot \partial_U f\left(X,U\right)\,,\\
U\cdot\nabla f\left(X,U\right)&= U\cdot\pl_X f\left(X,U\right)-\frac{U\cdot X}{X^2}\left(X\cdot\pl_X-U\cdot\pl_U\right)f\left(X,U\right)\,,\\
\nabla^2f\left(X,U\right)&=\partial^2_X f\left(X,U\right)-\frac1{X^2}\Big[(X\cdot\pl_X+d)X\cdot\pl_X-U\cdot\pl_U\Big]f\left(X,U\right)\\
& \hspace*{2cm}+2\,\frac{U\cdot X}{X^2}\,\pl_U\cdot\pl_{X}f\left(X,U\right)+\left(\frac{U\cdot X}{X^2}\right)^2\,\pl_U^2f\left(X,U\right)\,,
\end{align}
\end{subequations}
where $f\left(X,U\right)$ is a generic field in ambient space that is tangent \eqref{tangent}.

It is furthermore often convenient to adopt the notation \eqref{fnxu}
\begin{equation}
    f^{(n)}(X,U)\equiv\frac1{(X^2)^n}\,f(X,U)\,.
\end{equation}
The ambient field $\varphi^{(n)}_s$ has gauge transformation:
\begin{align}
\delta_{\xi_{s-1}}\varphi^{(n)}_s\left(X,U\right)=U\cdot\pl_X\,\xi^{(n)}_{s-1}\left(X,U\right)+U\cdot X\,(d-2+2n+2\,U\cdot\pl_U)\xi^{(n+1)}_{s-1}\left(X,U\right)\,,
\end{align}
and homogeneity condition:
\begin{align}
(X\cdot\pl_X+\mu+2n)\varphi^{(n)}_s\left(X,U\right)=0.
\end{align}

Fields $\varphi^{(n)}_s$, $\varphi^{\prime (n)}_s$ and $\xi^{(n)}_{s-1}$ enjoy the identities:
\begin{subequations}\label{idmanipn}
\begin{align}
    \Box\varphi^{(n)}_s&=-2\,U\cdot X\,\pl_U\cdot\pl_X\,\varphi^{(n+1)}_s-U^2\,\varphi^{\prime(n+1)}_s\\ \nonumber &\hspace*{4cm}+2n(d+2n-4+2\,U\cdot\pl_U)\varphi^{(n+1)}_s\,,\\
    \pl_U\cdot\pl_X\,\varphi^{(n)}_s&=\frac12\left[U\cdot\pl_X\,\varphi^{\prime(n)}_s+U\cdot X\,(d-2+2n+2\,U\cdot\pl_U)\varphi^{\prime(n+1)}_s\right],\\
     \Box\varphi^{\prime(n)}_s&=-2\,U\cdot X\,\pl_U\cdot\pl_X\,\varphi^{\prime(n+1)}_s\\\nonumber&\hspace{2cm}
     -2(d+2U\cdot\pl_U)\varphi^{\prime(n+1)}_s+2n(d+2n+2\,U\cdot\pl_U)\varphi^{\prime(n+1)}_s\,,\\
     \Box\xi^{(n)}_{s-1}&=-2\,U\cdot X\,\pl_U\cdot\pl_X\,\xi^{(n+1)}_{s-1}\\\nonumber&\hspace{1cm}-2(d-2+2\,U\cdot\pl_U)\xi^{(n+1)}_{s-1}+2n(d+2n-4+2\,U\cdot\pl_U)\xi^{(n+1)}_{s-1}\,.
\end{align}
\end{subequations}

\subsection{Traceless ambient fields}
\label{app::opalg4}

Here we give the operator algebra for the null auxiliary ambient vector $W^2=0$ subject to the tangentiality constraint $X \cdot W=0$. We furthermore restrict to the hyperboloid $X^2+1=0$. 

\begin{subequations}
\begin{align}
\big[\nabla \cdot {\tilde D}_W,W\cdot\nabla \big]&=\bigg(\frac{d-1}{2}+W \cdot \partial_W\bigg)\nabla^2 \\ \nonumber & \hspace*{1cm}-\bigg( (W \cdot \partial_W)^2+\frac{3(d-1)}{2}\,W \cdot \partial_W +\frac{(d-1)^2}{2}\bigg)W \cdot \partial_W\,,
\\
\big[\nabla^2,W\cdot\nabla \big]&=-2 \big(\tfrac{d}{2}-1+W \cdot \partial_W\big)\,W\cdot \nabla \,,
\end{align}
\end{subequations}
where we introduced:
\begin{equation}
    {\tilde D}_W := \left(d-1+2 W \cdot {\cal P} \cdot  \partial_W\right) {\hat D}_W.
\end{equation}
From which one can derive:
\begin{subequations}
\begin{align}
[\nabla^2,(W\cdot\nabla)^n]f_{s-n}\left(X,W\right)&=-n(d-1+2s-n)\,(W\cdot\nabla)^nf_{s-n}\,,\\
[\nabla\cdot{\tilde D}_W,(W\cdot\nabla)^n]f_{s-n}\left(X,W\right)&=\tfrac{n}2\left(d-2+2s-n\right)(W\cdot\nabla)^{n-1}\\ \nonumber
& \hspace*{0.75cm}\times \left[\Box-(s+d-1)(s-1)+s-n\right]f_{s-n}\left(X,W\right)\,,
\end{align}
\end{subequations}
with $f_{s-n}$ a homogeneous function of degree $s-n$ in $W$.

\section{AdS Harmonic functions}
\label{app::adsharm}

Square integrable functions $F\left(x_1,u_1;x_2,u_2\right)$ on AdS$_{d+1}$ which depend only on the geodesic distance between $x_1$ and $x_2$ can be expanded in a basis of regular totally symmetric eigenfunctions $\Omega_{\nu,J}\left(x_1,x_2\right)$ of the AdS Laplacian:
\begin{subequations}
\begin{align}
\left(\Box_{1}^{2}+\left(\tfrac{d}{2}\right)^2+\nu^2+J\right)\Omega_{\nu,J}(x_1,x_2) & =0\,,
\end{align}
\end{subequations}
which are traceless and divergenceless:
\begin{equation}
\left(\nabla_1 \cdot {\hat \partial}_{w_1}\right)\Omega_{\nu,J}(x_1,w_1;x_2,w_2)=0.
\end{equation}
They satisfy the completeness relation:
\begin{equation}
\sum_{l=0}^s\int^\infty_{-\infty} d\nu \,c^{(s)}_{l}(\nu) (W_1\cdot\nabla_1)^l(W_2\cdot\nabla_2)^{l}\Omega_{\nu,s-l}(X_1,W_1;X_2,W_2) = 
\delta(X_1,X_2) (W_{1}\cdot W_2)^s\,,
\label{eq:Completeness}
\end{equation}
where
\begin{equation}\label{csl}
c^{(s)}_l=\frac{2^l (s-l+1)_l \left(\frac{d}{2}-l+s-\frac{1}{2}\right)_l}{l! (d-2 l+2 s-1)_l \left(\frac{d}{2}-l+s-i \nu \right)_l \left(\frac{d}{2}-l+s+i \nu \right)_l}\,,
\end{equation}
and the orthogonality relation:
\begin{equation}
 \int_{\text{AdS}} 
 dX\, \Omega_{\bar{\nu},s}(X_1,W_1;X,{\hat D}_W)
 \!\int^\infty_{-\infty}
 d\nu \,\Omega_{\nu,s}(X,W;X_2,W_2) 
 =
 \Omega_{\bar{\nu},s}(X_1,W_1;X_2,W_2)\,.
\end{equation}
They also admit the integral form \cite{Leonhardt:2003qu}
\begin{equation}
\Omega_{\nu,s}(X_1,W_1;X_2,W_2) = \frac{\nu^2}{\pi} \int_{\partial \text{AdS}}dP\, K_{\frac{d}{2}+i\nu,s}(X_1,W_1;P,{\hat D}_Z)K_{\frac{d}{2}-i\nu,s}(X_2,W_2;P,Z),
\end{equation}
where the $K_{\frac{d}{2}\pm i\nu,s}$ are the bulk-to-boundary propagators \eqref{bubospinsgen}.

In expressing the bulk-to-bulk propagators in this work in a basis of the above harmonic functions, it is useful to employ the following identities: 
\begin{multline}\label{divgradmarm}
\nabla_1\cdot\hat{D}_{W_1}\,\left[(W_1\cdot\nabla_1)^l\Omega_{\nu,s}(X_1,W_1;X_2,W_2)\right]\\=-\frac{l (d+l+2 s-2)}{2 \left(\frac{d-1}{2}+l+s-1\right)}\,\left[\left(\frac{d}{2}+l+s-1\right)^2+\nu ^2\right]\,(W_1\cdot\nabla_1)^{l-1}\Omega_{\nu,s}(X_1,W_1;X_2,W_2),
\end{multline}
\begin{multline}\label{boxgradharm}
\Box_1\,\left[(w_1\cdot\nabla_1)^l\Omega_{\nu,s}(X_1,W_1;X_2,W_2)\right]\\=-\left[l (d+l+2 s-1)+\left(\frac{d}{2}+i \nu \right) \left(\frac{d}{2}-i \nu \right)+s\right]\,(w_1\cdot\nabla_1)^{l}\Omega_{\nu,s}(X_1,W_1;X_2,W_2)\,.
\end{multline}

\end{appendix}

\bibliography{refs}
\bibliographystyle{JHEP}

\end{document}